\documentclass[%
 reprint,
 amsmath,amssymb,
 aps,nofootinbib, 
]{revtex4-1}
\usepackage{bm}
\PassOptionsToPackage{linktocpage}{hyperref}
\usepackage[hyperindex,breaklinks]{hyperref}
\usepackage{enumitem}
\usepackage{slashed}
\usepackage{comment}
\usepackage[dvipsnames]{xcolor}
\usepackage{tabularx}
\usepackage{array}
\usepackage{orcidlink}

\newcolumntype{x}[1]{>{\centering\arraybackslash\hspace{0pt}}p{#1}}

\renewcommand{\vec}[1]{\ensuremath{\boldsymbol{#1}}}

\newcommand{\iu}{\mathrm{i}}	

\usepackage{array}
\usepackage{mathtools}

\usepackage{etoolbox}

\begin{document}

\title{Kaluza-Klein Spectroscopy from Neutron Oscillations \\
into Hidden Dimensions }



\author{Gia Dvali, Manuel Ettengruber, Anja Stuhlfauth\,\orcidlink{0009-0005-0920-379X}  
} 
\affiliation{Arnold Sommerfeld Center, Ludwig-Maximilians-Universit\"at, Theresienstra{\ss}e 37, 80333 M\"unchen, Germany
}
\affiliation{Max-Planck-Institut f\"ur Physik, F\"ohringer Ring 6, 80805 M\"unchen, Germany
}

\date{\today}

\begin{abstract} 

Neutrons and neutrinos are natural probes for new physics.
Since they carry no conserved gauge quantum numbers, both can easily mix with the fermions from hidden sectors. 
A particularly interesting effect is the oscillation of a neutron or a neutrino into a fermion propagating in large extra dimensions.  
In fact, such a mixing has been identified as the possible origin of small neutrino mass. 
In this paper, we study neutron oscillations into an extra-dimensional fermion and show that this effect provides a resonance imaging of the Kaluza-Klein tower. 
The remarkable feature of this phenomenon is its generic nature:  because of a fine spacing of the Kaluza-Klein tower, neutrons 
at a variety of energy levels, both free or within nuclei,  find a bulk oscillation partner.
In particular, the partner can be a Kaluza-Klein mode of the same species that gives mass to the neutrino.  The existence of bulk states matching the neutron energy levels of nuclear spectra gives rise to tight constraints as well as to potentially observable effects. 
For a free neutron, we predict recurrent resonant oscillations occurring with the values of the magnetic field correlated with the KK levels. 
We derive bounds on extra dimensions from ultra-cold neutron experiments and suggest  signatures for refined measurements,  
which, in particular,  can probe the  parameter space motivated by  the Hierarchy Problem. 
Ultra-cold neutron experiments offer a unique way of Kaluza-Klein spectroscopy.

\end{abstract}

\maketitle

\section{Introduction}

The framework of large extra dimensions~\cite{Arkani-Hamed:1998jmv, Arkani-Hamed:1998sfv} (hereafter referred to as ADD model) is motivated by the solution to the Hierarchy Problem, the inexplicable smallness of the weak interaction scale relative to the Planck mass, $M_P$. 
In this theory, the fundamental cutoff of gravity, $M_f$, defined as the Planck mass of $4+N$-dimensional theory, is lowered relative to the four-dimensional Planck scale $M_P$. 
Correspondingly, all UV-sensitivities, including the one of the Higgs mass, are cut off at the scale  $M_f \ll M_P$.  
The lowering of the cutoff can be understood in different languages. 
 
First, this phenomenon has a clear geometric meaning. 
At separation $r$ smaller than the compactification radius $R$, the gravitational flux of point-like sources spreads according to high-dimensional Gauss's law.
Correspondingly, it is diluted in the large volume of extra-dimensional space, usually referred to as ``bulk".  
Because of this dilution, the strength of four-dimensional gravity that operates at large distances, $r \gg R$,  is effectively weakened: four-dimensional Newton's constant is suppressed relative to the fundamental one by the $N$-dimensional volume of the extra space,  $V_N$.     
The resulting relation between the high-dimensional and four-dimensional Planck scales is
\begin{equation} 
    M_f = \frac{M_P}{\sqrt{M_f^N V_N}}.
    \label{masterequation}
\end{equation}
For definiteness, we shall assume compactification on an $N$-torus of radii $R_1,R_2 \dots R_N$, with corresponding volume $V_N = (2 \pi)^N R_1 \dots R_N$.

An alternative interpretation of (\ref{masterequation}) was offered in~\cite{Dvali:2007hz,Dvali:2007wp,Dvali:2009ne}, where it was pointed out that (\ref{masterequation}) represents a manifestation of the following general relation, 
 \begin{equation}
    M_f =  \frac{M_P}{\sqrt{N_{sp}}}, 
    \label{species}
\end{equation}
where $N_{sp}$ is the number of particle species.
This expression tells us that in a $3+1$-dimensional theory with $N_{sp}$ particle species, the fundamental scale of gravity is lowered to the species scale given by (\ref{species}). 
Indeed, noticing that  $N_{sp} =  M_f^N V_N$ is the number of Kaluza-Klein (KK) species, it is clear that (\ref{species}) reproduces relation (\ref{masterequation}). 
Thus, the ADD model can be viewed as a geometric realization of the species effect (\ref{species}), describing it as the suppression by the volume.   

In the ADD framework, all the  Standard Model (SM)  species are localized on a brane with a $4$-dimensional world volume. 
The localization of the gauge fields, such as the photon~\cite{Dvali:1996xe}, imposes the condition that no charged particles (including the hypothetical ones beyond SM) can freely propagate in the bulk.

The volume suppression feature is shared by the interactions between the Standard Model particles and any other hypothetical bulk species, i.e., the species that propagate in the extra space.  
This feature has several interesting implications for the SM particles that carry no conserved gauge charges.
The two prominent low-energy candidates are the neutron~\cite{Dvali:1999gf} and the neutrino~\cite{Arkani-Hamed:1998wuz,Dvali:1999cn,Dienes:1998sb}. 
The feature of gauge neutrality allows both of these particles to serve as sort of ``portals" to hidden dimensions through their mixing with bulk degrees of freedom.  
  
Such mixing can shed new light on some of the long-standing puzzles of the Standard Model, e.g., such as the origin of the neutrino mass. 
A solution to the latter puzzle was proposed in~\cite{Arkani-Hamed:1998wuz,Dvali:1999cn,Dienes:1998sb} where it was suggested that a mixing of the SM  (active) neutrino with a bulk sterile fermion can be the origin of the neutrino's small mass.
Just as in the case of gravity, the smallness of the neutrino mass is the result of the volume suppression.
This mechanism of neutrino mass generation is accompanied by new phenomenological signatures, such as the possibility of oscillations of SM active neutrinos into the KK tower of the sterile bulk fermion~\cite{Dvali:1999cn}.
  
The possibility of a neutron portal into hidden dimensions was first pointed out in~\cite{Dvali:1999gf}.
As discussed in this work, neutron transitions into a hidden state are a generic feature of the brane-world scenario.
This is due to the transportation of neutrons across the bulk by virtual brane bubbles, so-called ``baby branes". 
In such a process, a neutron can tunnel across the bulk to a nearby parallel brane. 
   
For an observer inhabiting the world-volume of our brane, the process will effectively be described as a transition of a neutron into a hidden particle.
This process is non-perturbative in its origin.
Its probability is exponentially suppressed but never zero, as it is permitted by all conservation laws.
The suppressed rate makes it naturally compatible with the phenomenological bounds on neutron disappearance.  
   
In the present paper, we shall investigate a different process of neutron transitions into extra dimensions.  Namely, via its mixing with a bulk fermion. 
This process is interesting because of the following specifics. 
First, again, the mixing between neutrons and bulk sterile fermions is expected to be rather generic, since none of them carry any conserved SM gauge quantum numbers. 
 
Secondly, there is a qualitative novelty with respect to theories in which the neutron mixes with a partner of a fixed mass, such as the neutron from a mirror SM \cite{Berezhiani:2005hv} or 
many hidden copies of the SM \cite{Dvali:2009ne}.
 In such theories, due to a single available partner, the resonance transition requires 
coincidences that are difficult to control. 
This difficulty is most prominent in the case of a single mirror copy. 
The mirror symmetry is inevitably broken by environmental factors such as the nuclear binding energy of the neutron, or an ambient magnetic field.
As a result, the transition becomes unobservably suppressed, unless one assumes certain coincidences and/or cancellations of external factors, such as, for example, the magnetic field in the hidden sector~\cite{Berezhiani:2009ldq}. 
 
The transition of a neutron into extra dimensions offers a qualitatively different picture. 
The reason is a finely spaced tower of KK states in the ADD model. 
Because of the high density of  KK states, neutrons in a wide spectrum of energy levels 
find a nearly degenerate bulk partner into which they can oscillate. 
 A variety of bound or free neutrons can become portals into extra space.
This puts severe constraints on the parameters of the theory and at the same time provides new signatures for experimental searches. 

A particularly interesting signature comes from the recurrent resonant oscillations of a free neutron into the KK modes, taking place 
for a sequence of special values of the magnetic field.
Unlike the scenarios with a single oscillation partner with the fixed mass, in the case of extra dimensions, the resonant transitions take place for the quantized values of the magnetic field matching the KK spectrum. 
In this way, the neutron provides a magnetic imaging of the KK tower.

Because of such effects, the scenario is subjected to non-trivial bounds from current experiments with ultra-cold neutrons \cite{nEDM:2020ekj, Ban:2023cja}. Fitting the data of these experiments, we derive bounds on the parameters of extra dimensions. These bounds are already probing the parameter regimes motivated by the Hierarchy Problem.
We also discuss new signatures for future refined measurements.

\section{General Framework}

In this section, we review the basics of the ADD model~\cite{Arkani-Hamed:1998jmv,Arkani-Hamed:1998sfv}, preparing the basis for further analysis. 
An intrinsic property of extra dimensions is the existence of KK modes. 
Each bulk particle gives rise to a tower of four-dimensional massive modes. 
That means that each extra momentum eigenstate, from a $3+1$-dimensional perspective, corresponds to an independent massive particle.
Because of the periodicity of extra coordinates, the momenta (and correspondingly the $4$-dimensional masses) are quantized. 
For $N$ extra dimensions, the masses of the KK modes $m_{\vec{k}}$ are labeled by a set of integers that can be represented as $N$-component vector $\vec{k}= (k_1, \dots, k_N)$ and are given by
\begin{equation}
    m_{\vec{k}} \, = \, \sqrt{\frac{k_1^2}{R_1^2} + \dots + \frac{k_N^2}{R_N^2}} \,.
    \label{masssplitting}
\end{equation}
The counting holds for each bulk field independently. 
It is therefore clear that a single bulk field (e.g., the graviton) gives rise to $N_{\rm KK} =  M_f^N V_N$ four-dimensional species.

Let us review some constraints on ADD parameters.
These come from various phenomenological, astrophysical, and cosmological considerations, and have been discussed already in~\cite{Arkani-Hamed:1998sfv}.
More updated versions of these constraints represent the refinements, in particular, in the light of the improved experimental data. 
Among the phenomenological constraints, the most important ones are the following two model-independent bounds. 

The first is the bound on the fundamental scale, 
\begin{equation}
    M_f \,  \gtrsim 10 {\rm TeV} \,, 
    \label{LHC}
\end{equation}
imposed by the LHC experiments ATLAS~\cite{ATLAS:2021kxv} and CMS~\cite{CMS:2021far}, due to non-observation of new particle resonances at the energy scale probed by this collider.  
ADD predicts the existence of such resonances since the scattering at momentum transfer $M_f$ saturates perturbative unitarity.
The unitarization requires the presence of new resonances.
They shall come in the form of the tower of quantum microscopic black holes and/or string resonances~\cite{Antoniadis:1998ig}. 

The second model-independent constraint comes from the modification of Newton's inverse square law.
An immediate consequence of ADD theory is that Newton’s inverse square law is modified at distances shorter than the radii of the extra dimensions. 
For example, taking all radii equal to $R$, at a separation $r \ll R$, Newton's law changes into 
\begin{equation}
  \frac{1}{M_P^2}\frac{1}{r}   \, ~ \,  \rightarrow \, ~ \,  \frac{1}{M_f^{2+N}}\frac{1}{r^{1+N}} \,. 
    \label{Newton}
\end{equation}
This modification can be understood as the result of the exchange of the KK tower of gravitons, 
\begin{equation}
 \lim_{R \rightarrow \infty}  \frac{1}{M_P^2} \sum_k \frac{e^{-m_kr}}{r}  \, = \,   \frac{1}{M_f^{2+N}}\frac{1}{r^{1+N}} \,, 
    \label{NewtonLimit}
\end{equation}
where masses of KK gravitons $m_k$ are given by (\ref{masssplitting}). 
Each graviton mediates a Yukawa interaction and the summation over them gives a power-law high-dimensional Newtonian potential. 
 
The physical meaning of the above equation is easy to understand. 
First notice that, because of the exponential suppression of modes with $m_k > 1/r$, at each separation only the modes with $m_k < 1/r$ contribute. 
For them, the exponential  e$^{-m_kr}/r$ can be replaced by $1/r$.  
Thus, at any scale, each light mode increases Newton's force by one unit.  
At the same time, the multiplicity of such modes, for  $r \ll R$ is $\simeq (2\pi R/r)^N$ and the potential can be approximated as 
\begin{align}
    \frac{1}{M_P^2} \sum_k &\frac{e^{-m_kr}}{r}  \, \simeq \,  \frac{(2\pi R)^N}{M_P^2} \frac{1}{r^{1+N}} \, = \, \\ \nonumber
    =& \frac{1}{M_f^{2 +N}} \frac{1}{r^{1+N}} \,. 
    \label{NewtonLimit1}
\end{align}
 which is simply the $4+N$-dimensional Newtonian potential with the fundamental scale $M_f$ that is related to $M_P$ via (\ref{masterequation}).

In the continuum limit $R \rightarrow \infty$, the equality becomes exact. 
Notice that in the continuum limit the summation over a discrete KK tower becomes an integration over the continuum of extra-dimensional momenta $\vec{p} = \frac{\vec{k}}{R}$, 
\begin{equation}
    \lim_{R \rightarrow \infty}  \sum_k   =   V_N \int \frac{d^N \vec{p}}{(2\pi)^N}  \,.
    \label{summ}
\end{equation}
From here it also follows (see the appendix) that the multiplicity of the  KK modes close to a given mass $m \gg 1/R$ is,
\begin{equation}
    Z \,  \sim   \, (2\pi Rm)^{N-1} \,. 
    \label{DEG}
\end{equation}
Non-observation of any modification of Newton's law in the tabletop experiments puts the following upper bound on the radius of the largest extra dimension,~\cite{ParticleDataGroup:2022pth,Lee:2020zjt, Adelberger:2009zz, Tan:2016vwu}, 
\begin{equation}
    R_{max} \, \lesssim \,  30 \,  \mu {\rm m} \,. 
    \label{RRR}
\end{equation}

On top of these constraints there exist others that emerge based on further assumptions about the parameters.
For instance, as shown in~\cite{Arkani-Hamed:1998sfv}, 
the star-cooling bound can alter the value of  $M_f$ or of the radii, depending on $N$ and assuming that all radii are equal. 
However, as shown in~\cite{Dvali:1999cn} this alteration is avoided if some of the radii are smaller.
In this case, the supernova cooling constraints are weakened and the bound on the largest extra dimension (\ref{RRR}) and on $M_f$(\ref{LHC}) are unaffected.

Similarly, the decay of relic KK gravitons produced either cosmologically or via supernovae cooling, can alter the diffuse gamma ray spectrum.
However, as explained in~\cite{Arkani-Hamed:1998sfv}, the diffuse cosmic gamma-ray bound is rather sensitive to the bulk physics 
and can easily be avoided, in particular,  in cases in which KK gravitons have other decay channels into the bulk species (e.g., the ones inhabiting ``fat" branes). 

The diffuse gamma-ray bound was later reconsidered in~\cite{Hannestad:2001jv} by taking into account an additional contribution coming from KK gravitons produced in supernova stars throughout the history of the Universe and was concluded to be more severe.
However, these authors put aside the loopholes pointed out 
in the previous analyses of~\cite{Arkani-Hamed:1998sfv} which significantly weaken the bounds. 
Thereby, their refinements fall in the category that is strongly sensitive to the details of the model, such as the shape of the compact manifold and other unknowns of bulk physics. 

There exists an extensive literature with several other important model-dependent constraints, which we shall not discuss here due to lack of space.
These must be taken into account within specific realizations of the ADD framework. 
However, the bottom line is that, currently, with no further assumptions, the universal constraints are (\ref{LHC}) and (\ref{RRR}).

\section{Neutrino oscillations into large dimensions} 
 
Before we move to the discussion of the neutron, we wish to review the oscillations of neutrinos into hidden dimensions studied in~\cite{Dvali:1999cn}.
This analysis was done within the theory of neutrino mass introduced in~\cite{Arkani-Hamed:1998wuz}.  
In this paper, it was proposed that the ADD setup provides a natural framework for generating small neutrino masses via their mixing with bulk sterile partners.
For definiteness, we shall consider a single active SM neutrino species and correspondingly a single species of a bulk sterile neutrino.   
We denote this bulk field by $\Psi$.
 
The field $\Psi(x,y)$ is a function of four space-time coordinates $x$ and $N$ extra space coordinates $y$.
It transforms in the spinor representation of the $4+N$-dimensional Poincare symmetry.
For simplicity, we shall choose it as the smallest representation of this sort. 
However, at a more fundamental level, other considerations, e.g., the cancellation of gravitational anomalies, may apply.

From the point of view of the four-dimensional Lorentz group acting on space-time $x$, the bulk fermion $\Psi$ contains both chiralities.  
Moreover, $4+N$-dimensional chirality is only defined in even space-time dimensions, which in the present case implies $N=2n$ with $n$ an integer.
There exists no chirality for $N=2n-1$.
The dimensionalities of the irreducible massless spinors in $N=2n$ and $N=2n-1$ are the same and equal to $2^{2+n-1}$.

The reduction of a generic $4+N$-dimensional spinor into the irreducible representations of ($4D$-Lorentz)$\times SO(N)$ symmetry has the following schematic form, 
\begin{equation} \label{REDUCTION}
    2^{2+n -1}  \,  \rightarrow  \,   2_{L} \times 2^{n-1}   + 2_{R} \times {2^{n-1}}  \, ,
\end{equation}
where numbers indicate the dimensionalities of the representations and only the $4$-dimensional chiralities are labeled explicitly by $L, R$.  The chiralities with respect to the internal $SO(N)$-symmetry depend on the type of the initial spinor, as discussed above.  
  
Thus, a massless bulk fermion $\Psi$ under the $4$-dimensional Lorentz symmetry decomposes into $2^{n-1}$ left-handed and $2^{n-1}$ right-handed fermions: 
\begin{equation} \label{PSIA}
    \Psi \,  \rightarrow  \,   \sum_{A} \Psi_{ L}^{(A)}\:  \,   +   \,  \sum_{\bar{A}}   \Psi_{R}^{(\bar{A})} \,,     
\end{equation}
where  $A, \bar{A}$ label the basic $SO(N)$-spinors 
\footnote{That is, indices $A, \bar{A}$ label the types of spinors forming a proper complete basis in the $SO(N)$ spinor space, not to be confused with a spinor index.}.

The above decomposition takes place for each KK level separately,   
\begin{equation} \label{PSIKK} 
    \Psi(x,y) \, \rightarrow \,  \frac{1}{\sqrt{V_N}} \sum_{\vec{k}} e^{\frac{i\vec{k}\vec{y}}{R}} \left ( \sum_{A} \Psi_{\vec{k}, L}^{(A)}(x) \:  \,   +   \,  \sum_{\bar{A}}   \Psi_{\vec{k}, R}^{(\bar{A})}(x)  \right )\,. 
\end{equation}  
Notice that a canonically normalized bulk fermion $\Psi$ has mass dimension $\frac{3+N}{2}$, whereas the KK modes $\Psi_{\vec{k}}$ have mass dimension $\frac{3}{2}$.

Upon dimensional reduction from $4+N$ to $4$-dimensional space-time, the extra-dimensional part of the Dirac operator, $\bar{\Psi} \vec{\gamma} \vec{\partial}_y \psi$, provides the Dirac mass terms for the KK modes. 
In this process, at each KK level,  the left and right chiralities pair up and produce the tower of massive Dirac fermions: 
\begin{equation} \label{PSIKKMASS} 
    \sum_{\vec{k}}  m_{\vec{k}}  \sum_{A= \bar{A}=1}^{2^{n-1}} \bar{\Psi}_{\vec{k}, L}^{(A)}(x) \Psi_{\vec{k}, R}^{(\bar{A})}(x) \,. 
\end{equation}  
The labeling $A= \bar{A}$ is assumed to be such that the spinors get filtered through the high-dimensional $\vec{\gamma}$-matrix properly. 
That is, each KK level  $\vec{k} \neq 0$ gives rise to $2^{n-1}$ massive Dirac fermions.    
     
The fermions corresponding to the KK level  $\vec{k} \, = \, 0$ remain massless.  
A certain superposition of their right-handed components plays the role of the right-handed partner of the SM neutrino, which endows the latter with a Dirac mass.
Let us discuss this effect in more detail.           
   
The  mass term for the neutrino arises from a Yukawa-type interaction involving the left-handed lepton doublet $L = (\nu_L, e_L)$, the bulk fermion $\Psi$ and the Higgs doublet field $H = (H^0, H^-)$,
\begin{equation}\label{eq:neutrinodiracExtraDim1}
    \mathcal{L}_{\rm int} \, = \, \frac{1}{M_*^{N/2}} H(x) \, \bar{L}(x) \, \Psi(x, y=0)\,.
\end{equation}
The brane is localized at $y=0$.  
The above expression is rather schematic and requires further specification. 
First, the requirement imposed by the consistency of the four-dimensional effective field theory is that it must be invariant under the 
four-dimensional Poincare symmetry as well as the gauge symmetry of the SM.    

The invariance of (\ref{eq:neutrinodiracExtraDim1}) under the SM gauge symmetry uniquely fixes $\Psi$ to be a gauge-neutral degree of freedom. 

At the same time, the four-dimensional Poincare invariance implies that only the right-handed components $\Psi_{\vec{k}, R}^{(\bar{A})}(x)$  (or the charge conjugates of the left-handed ones) of the bulk fermion $\Psi$ participate in the above Yukawa coupling. 
However, this requirement does not fully specify the form of the coupling because of the following reason. 
 
Since the brane breaks the higher-dimensional Poincare symmetry, it has no ``obligation" to respect its $SO(N)$ subgroup.        
That is, in general, the brane acts as a spurion that absorbs the $SO(N)$-spinor index.
The pattern of this breaking defines to which superposition of the bulk spinors $\Psi_{R}^{(\bar{A})}$ the SM neutrino couples. 
      
Within the effective field theory, this information is not available.
It depends on the origin of the brane and on the dynamics that localize the $4$-dimensional chiral fermions on it.
In our phenomenological study, we shall treat this as an input.
Note that the smallness of the mass of the SM neutrino is largely insensitive to the precise form of the $\Psi_{R}^{(\bar{A})}$-superposition to which it mixes since the mass comes predominantly from the $\vec{k}=0$ mode.
However, the mixing with higher energy states in the KK tower will depend on it.

Upon taking into account the Higgs vacuum expectation value $\langle H \rangle = (v,0)$, the above coupling reduces to a mass term that mixes the active left-handed neutrino, with the tower of KK modes,
\begin{equation}\label{eq:neutrinodiracExtraDim}
    \mathcal{L}_{\rm int} \, = \,  \alpha_{\nu}  \, \bar{\nu}_L(x) \, \sum_{\vec{k}}  \nu_{k, R} \,,
\end{equation}
where 
  \begin{equation}
   \alpha_{\nu} \, \equiv \,  \frac{v}{\sqrt{M_*^N V_N}}  \,,
\end{equation}
and 
   \begin{equation} \label{CPSI} 
   \nu_{k, R} \,  \equiv \,  \sum_{\bar{A}}  c_{\bar{A}}\Psi_{\vec{k}, R}^{(\bar{A})}(x) 
   \end{equation}
denotes the superposition of spinors from each KK level to which $\nu_L$ mixes. 
In general, the coefficients $c_A$ can also depend on the level $k$.
We shall make a simplified assumption that the $\nu_{k, R}$ are the eigenstates of the KK masses.
In this case, the orthogonal superpositions will effectively decouple from our problem and can be ignored. 
 
Mixing of $\nu_L$ with the tower of $\nu_{k, R}$ generates the mass of neutrino, $m_{\nu} \simeq \alpha_{\nu}  \sqrt{\sum_{\bar{A}} |c_{\bar{A}}|^2}  \sim \alpha_{\nu}$, which for $M_* \sim  M_f$ is
\begin{equation}
    m_\nu \sim \frac{v M_f}{M_P} \,.
\end{equation}
For $M_f \sim 10$TeV, this value is in the right phenomenological ballpark.  
 
The mass of the neutrino is generated predominantly through the mixing with the right-handed components of the $\vec{k} =0$ level, which have no masses of their own.
However, mixing with higher members of the KK tower results in oscillations of the neutrino into these states.
This can have potentially observable effects~\cite{Arkani-Hamed:1998wuz,Dvali:1999cn}.

For simplicity, let us reproduce this effect for one relevant extra dimension with radius $R$.
In this analysis, we shall closely follow~\cite{Dvali:1999cn}. 
In this case, the indices $A, \bar{A}$ are not required and we can write,  
\begin{equation}
    \Psi(x,y) = \frac{1}{\sqrt{2\pi R}} \sum_{k = - \infty}^{k = \infty}  e^{\frac{iky}{R}} \left (\Psi_{k, L}(x) \:  \,   +   \, \Psi_{k, R}(x)  \right )\,. 
\end{equation}
Notice that the hierarchy $M_f/M_P \sim 10^{15}$ is still maintained by assuming the existence of additional $N-1$ dimensions of much smaller radii. 
The corresponding KK excitations are heavy and effectively decouple from the neutrino mass problem.  

Because of the degeneracy under the reflection $k \rightarrow -k$ for each $k \neq 0$ level, the active neutrino mixes with the modes $\nu_{k R} \equiv \frac{1}{\sqrt{2}} (\psi_{k,R} + \psi_{-k,R})$, whereas the orthogonal superpositions, $\frac{1}{\sqrt{2}} (\psi_{k,R} - \psi_{-k,R})$, decouple and can be neglected.

Then, the part of the Lagrangian describing the relevant mass terms is
\begin{equation}\label{eq:neutrinodirac}
    \mathcal{M}_{\nu} =  \alpha_{\nu}  \nu_L  \left (\nu_{0 R}   \, + \,   \sqrt{2} \sum_{k=1}^{\infty}   \nu_{k R} \right ) + \sum_{k =1}^{\infty} \frac{k}{R} \bar{\nu}_{k L}  \nu_{k R} \,.
\end{equation}
Thus, we have the mass matrix of the following form, 
\begin{equation}
   \begin{pmatrix}
        0 & \alpha_{\nu} & \sqrt{2} \alpha_{\nu} & \sqrt{2} \alpha_{\nu} & ...\\
        0 & 0 & 0 & 0 & ...\\
        0 & 0 & 1/R & 0 & ...\\
        0 & 0 & 0 & 2/R  &... \\
        ... & ... & ... & ...  &... \\
    \end{pmatrix}.
\end{equation}
Each $k \neq 0$ mixes with SM neutrino with an angle given by $\tan \varphi_k = \frac{\alpha_{\nu} R}{k}$.
Thus, the modes with higher values of $k$ quickly decouple and the main effect is concentrated in the part of the KK tower with $k \sim 1$.  

The active (left-handed) SM neutrino represents a superposition of the mass eigenstates of the form
\begin{equation} \label{LActive} 
    \nu_L \, = \, \frac{1}{\mathcal{N}} \Big(\nu_L'  + \sum_{k= 1}^{\infty} \frac{\alpha_{\nu} R}{k} \nu_{k,L}' \Big),
\end{equation}
where $\mathcal{N}^2 = 1+ \sum_k \frac{(\alpha_{\nu} R)^2}{k^2}  \simeq 1$ is a normalization factor.
The mixing with the KK tower results in the oscillations of the active flavor into the KK modes.  
The survival probability is
\begin{equation}
    P_{\rm surv}(t) = |\langle \nu_L |\nu_L(t) \rangle|^2 = \frac{1}{\mathcal{N}^4} \Big| 1 + \sum_{k= 1}^{\infty} \frac{(\alpha_{\nu} R)^2}{k^2} \exp{(\iu \phi_k)} \Big|^2.
\end{equation}
Thus, we obtain an interference of the infinite number of oscillating modes with growing frequencies $ \propto k^2$ and decreasing amplitudes $ \propto 1/k^2$. 
Because of this, in practice, the higher-frequency modes can be averaged out and only a few low-frequency modes are observable. 
Some implications of this phenomenon, in particular for solar neutrinos, 
were discussed in~\cite{Dvali:1999cn}.
More updated experimental constraints can be found in 
\cite{Machado:2011jt} and in subsequent papers, with the latest bounds
appearing in \cite{Forero:2022skg}.

Notice that with the assumption of the alignment of the interaction (\ref{CPSI}) and mass (\ref{PSIKKMASS}) eigenstates, the above-reproduced analysis of~\cite{Dvali:1999cn} can be straightforwardly generalized to an arbitrary number of relevant dimensions.  
    
We shall now move to a discussion of an analogous story of neutron oscillations into extra dimensions.

\section{Neutron Oscillations into hidden dimensions}

The main novelty of the present paper is the possibility of neutron oscillations into extra dimensions.    
Similarly to the neutrino, the neutron carries no conserved gauge charge and can freely mix with the bulk species. 
We consider the oscillation of the neutron into KK modes of a bulk fermion.
As in the neutrino case, we shall denote this bulk fermion by  $\Psi$.
To start with, we assume that $\Psi$ is massless.  
    
We shall keep its origin generic.
However, a particularly motivated possibility would be the case in which $\Psi$ is the bulk partner of one of the SM neutrinos that generates its mass via the above-discussed mechanism.
Consequently, the neutron would mix with the same KK-tower as the active neutrino. 
Notice that such a mixing is totally safe from the neutrino mass perspective.
Because of a very large splitting between the masses of neutrino and neutron, they predominantly mix with the highly separated sectors of the KK spectrum.
Correspondingly, they have essentially zero influence on one another.
This is a rather economic scenario.
Nevertheless, we shall keep the discussion maximally general.

It is reasonable to assume that the mixing between the neutron and the bulk fermion originates from a more fundamental four-fermi interaction of the type, 
\begin{equation}\label{actionNPsi}
    \mathcal{S}_{\rm int} =   \int dx^4  \frac{1}{M_*^{2 + N/2}}  \overline{udd} \,  \Psi \,  + \, {\rm h.c.} \,, 
\end{equation}
where $M_*$ is a scale.
We shall not specify its origin and shall treat it as a phenomenological parameter. 
In general, it is reasonable to expect that $M_*$ is of order or higher than $M_f$. 

In the effective low-energy theory below the scale of the QCD confinement, the coupling (\ref{actionNPsi}) translates into an effective mass term that mixes the neutron with $\Psi$.
The effective Lagrangian is derived by the following replacement, 
\begin{equation}
    udd \rightarrow  \Lambda_{\rm QCD}^3\, n \,,
    \label{UUDN}
\end{equation}
where  $n$ is the neutron field and $\Lambda_{\rm QCD} $ is of the order of the QCD scale. 

As previously, we expand $\Psi$ into the KK modes (\ref{PSIKK}). 
The difference from the neutrino case is that the neutron is a Dirac fermion. 
Therefore, both chiralities, $n_L$, and $n_R$, are available for mixing with the respective components of the KK tower:  
\begin{eqnarray} 
    \label{PSILR}  
    \Psi_{\vec{k}, R}(x) \equiv  && \frac{1}{\sqrt{\sum_{\bar{A}}|c_{\bar{A}}|^2}} \sum_{\bar{A}}  c_{\bar{A}}\Psi_{\vec{k}, R}^{(\bar{A})}(x) \, ,  \nonumber \\
    \Psi_{\vec{k}, L}(x) \equiv  &&  \frac{1}{\sqrt{\sum_A |c_A|^2}} \sum_{A}  c_{A}\Psi_{\vec{k}, L}^{(A)}(x) \, . 
  \end{eqnarray}

We again make a simplified assumption that the combinations (\ref{PSILR}) with $A = \bar{A}$ are KK mass eigenstates.
That is, modulo mixing with the neutron, at each KK level $\vec{k}$ they form a four-dimensional Dirac fermion $\Psi_{\vec{k}} \, \equiv \,    \Psi_{\vec{k}, L}(x) \, + \,  \Psi_{\vec{k}, R}(x)$ with mass $m_{\vec{k}}$.    
In such a case, the orthogonal combinations decouple and we get the following effective $4$-dimensional Lagrangian describing the Dirac neutron, the KK tower of Dirac fermions $\Psi_{\vec{k}}$ and their mixing, 
\begin{equation}
\begin{split}
    \mathcal{L} =& \; \bar{n} \iu \slashed{\partial} n - m_n \bar{n} n \\
    &+\sum_k \Big( \bar{\Psi}_k \iu \slashed{\partial} \Psi_k - m_k \bar{\Psi}_k \Psi_k \Big) \\
    &+ \alpha \sum_k \bar{n} \Psi_k + h.c.,
\end{split}
\end{equation}
where $m_n$ is the SM mass of neutron and $\alpha$ is the mixing mass term, 
\begin{equation}\label{alpha}
\alpha \equiv   \frac{\Lambda_{\rm QCD}^3}{M_*^{2+N/2} \sqrt{V_N}} \,.
\end{equation}
We have absorbed the normalization factor  $\sqrt{\sum_A |c_A|^2}$  into the rescaling of parameters.  

Notice that $\alpha$ is an extremely small mass scale.  
In particular, taking into account that  $M_* \geqslant M_f$ and the current experimental bound (\ref{LHC}), we get the following upper bound on $\alpha$, 
\begin{equation}
    \label{alphabound}
    \alpha \lesssim 10^{-24} {\rm GeV} \,. 
\end{equation}
This quantity is much below any other scale in the problem. 
This allows us to study the oscillation picture perturbatively in $\alpha$.  

The masses and mixings between the neutron and the bulk states are described by the following mass matrix, 
 \begin{equation}
    \begin{pmatrix}
        m_n & \alpha & \alpha & \alpha\\
        \alpha & 0 & 0 & 0\\
        \alpha & 0 & m_{\vec{k}} & 0\\
        \alpha & 0 & 0 & m_{\vec{k'}}
    \end{pmatrix}.
\end{equation}
Notice that because of the bound (\ref{alphabound}), we have $\alpha/m_n \lesssim 10^{-24}$. 
In the leading approximation, each KK mixes with the neutron with an angle given by $\tan \varphi_k = \frac{\alpha}{\Delta m_k}$, where $\Delta m_{\vec{k}} = |m_n - m_{\vec{k}}|$. 
Putting aside miraculous coincidences, the smallest value of  $\Delta m_{\vec{k}}$ is set by the level splitting between the KK modes.
 
The picture is very similar to what we encountered in the neutrino case but with two differences. 
First, the mass matrix is symmetric.
Secondly, $\Delta m_{\vec{k}}$ reaches the minimum not at the bottom of the KK spectrum but in the region closest to the energy of the neutron.

Because of the mixing with the KK tower, the ordinary neutron, which is a SM interaction eigenstate, is not an exact mass eigenstate.
Labeling the mass eigenstates by primes, the SM neutron represents a superposition of the mass eigenstates of the following form, 
\begin{equation}
    n \, = \, \frac{1}{\mathcal{N}} \Big(n' + \sum_{\vec{k}} \frac{\alpha}{\Delta m_{\vec{k}}} \Psi_{\vec{k}}' \Big),
\end{equation}
where $\mathcal{N}^2 = 1+ \sum_k \frac{\alpha^2}{\Delta m_k^2} = 1 + \mathcal{O}(\alpha^2)$ is a normalization factor.
This state evolves in time as, 
\begin{equation}
    n(t) = \frac{1}{\mathcal{N}} \Big(n' + \sum_{\vec{k}} \frac{\alpha}{\Delta m_{\vec{k}}} e^{\iu \phi_{\vec{k}}} \Psi_{\vec{k}}' \Big),
\end{equation}
where $\phi_{\vec{k}} = |m_n - m_{\vec{k}}|t$.
The corresponding survival probability of the neutron is,
\begin{equation}
    P_{\rm surv}(t) = |\langle n|n(t) \rangle|^2 = \frac{1}{\mathcal{N}^4} \Big| 1 + \sum_{\vec{k}} \frac{\alpha^2}{\Delta m_{\vec{k}}^2} \exp{(\iu \phi_{\vec{k}})} \Big|^2.
\end{equation}

\begin{figure}[!ht]
    \centering
    \includegraphics[width=0.8\columnwidth]{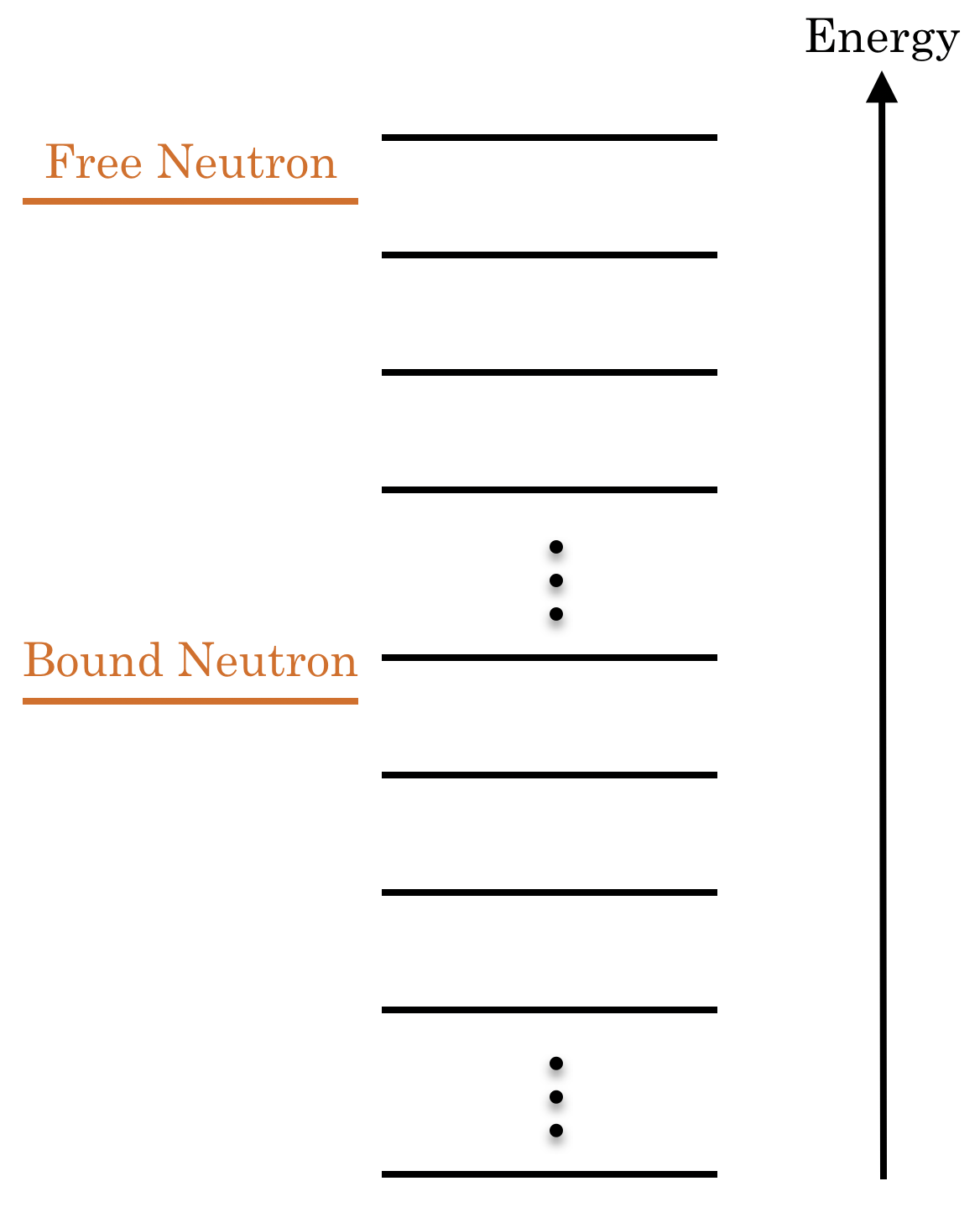}
    \caption{A schematic description of the matching of the energy levels of free and nuclear-bound neutrons with the KK spectrum of $\Psi$.
    }
    \label{fig:energylevels}
\end{figure}

Similarly to the neutrino case, the oscillation takes place in a collection of modes. 
The ``resonant"  modes correspond to KK levels that are closest to the SM energy of the neutron, $m_n$.
The more distant levels oscillate in increasing frequencies and suppressed amplitudes. 
Averaging over all modes except the level which is closest to $m_n$, we get the following expression for the probability of the neutron oscillating into $\Psi$,
\begin{equation}
    P_{n \to \Psi}(t) = \frac{Z}{\mathcal{N}^4} \frac{4 \alpha^2}{\Delta m^2} \sin^2 \Big( \frac{\Delta m}{2} t \Big).
    \label{survivalprobZ}
\end{equation}
 where $\Delta m \equiv  |m_n - m_{\vec{k}}|$ denotes the smallest mass splitting and the factor $Z$ accounts for the corresponding degeneracy.  
 For visualization, see Fig. \ref{fig:energylevels}.
 
 For a single extra dimension of radius $R$, we have $Z = 2$ because the mass is degenerate for $k$ and $-k$.
 Equivalently, the neutron mixes with the states $\Psi_{k} \equiv \frac{1}{\sqrt{2}} (\psi_{k} + \psi_{-k})$, whereas the orthogonal superpositions decouple.  
 The corresponding survival probability is 
 \begin{equation}
    P_{n \to \Psi}(t) = \frac{1}{\mathcal{N}^4} \frac{8 \alpha^2}{\Delta m^2} \sin^2 \Big( \frac{\Delta m}{2} t \Big)\,,
    \label{survivalprob}
\end{equation}
where $\Delta m \leqslant 1/R$.
Although ``accidentally"  $\Delta m$ may appear arbitrarily below this upper bound, a natural value would be $\Delta m \sim 1/R$.

For a higher number of relevant extra dimensions, the level-splitting is as small as $\Delta m \sim 1/(m_nR)^2$ (see the Appendix). 
The degeneracy count is the following. 
For  $N_R$ relevant dimensions of equal radii $R$ (again, $N_R$ must not be confused with the total number $N$ of large extra dimensions, $N_R \leqslant N$), in general, the number  $Z$ of KK states within a gap  $\Delta m$ 
satisfies, 
\begin{equation} 
    \label{Zm}
   \frac{Z}{\Delta m} \sim \frac{1}{m_n} (m_nR)^{N_R} \,.
\end{equation}

Taking this into account,  the survival probability for $N_R$ relevant dimensions of radius $R$ can be presented in the following form  
\begin{equation}
    P_{n \to \Psi}(t) \, \sim \,  \frac{m_n}{\Delta m} (m_n R)^{N_R} \left (\frac{\alpha}{m_n} \right )^2 \sin^2 \Big( \frac{\Delta m}{2} t \Big)\,.
    \label{Puseful}
\end{equation}
The oscillation amplitude is the largest for the states with the smallest $\Delta m$. 
Equipped with these equations, we shall next discuss the effect of neutron disappearance.

\section{Phenomenological bounds from neutron disappearance in nuclei}

In the previous section, we discussed how in the ADD framework a neutron can oscillate into particles propagating in extra dimensions.
We shall now find phenomenological constraints on such a process.
We first study the case of a massless bulk partner. 
In this case, the most severe constraints come from the non-observation of the disappearance of neutrons from the nuclei.

Let us consider an atomic nucleus with several neutrons in it.  
The above-derived formulas can be directly applied to such neutrons.
The mass $m_n$ must be understood as the energy level of a neutron in the given bound state.  
The neutron has a finite probability of transitioning into the bulk KK modes. 
From the point of view of the SM observer, this process is viewed as the disappearance of the neutron.
The vacated nucleus will get de-excited. 
Namely, a neutron from a higher level will occupy the state freed by the disappearing neutron.   
The process of de-excitation is accompanied by the emission of a hard photon, i.e., a photon of nuclear energy. 
      
For large extra dimensions, the typical oscillation time, set by the KK level splitting $\sim 1/\Delta m$, is much longer than the nuclear transition time.
In such a case, after the de-excitation, the process of coming back is no longer possible, since the place of the original neutron is occupied.    
Therefore, such a process would lead to a decay of atoms into isotopes with fewer neutrons.
This is not observed in nature and the experimental bound on the lifetime of neutrons within  nuclei is~\cite{KamLAND:2005pen}
\begin{equation} \label{taun}
    \tau_n \, > \, 10^{30} \, {\rm y} \, \sim \,  10^{62} \, {\rm GeV}^{-1} \,. 
\end{equation}

This bound translates to a bound on the neutron disappearance rate in our scenario.  
The rate can be derived from (\ref{survivalprob}) as the average transition probability  per unit time, 
\begin{equation}
    \lambda_n \, 
        = \,  \frac{2 Z \alpha^2}{\Delta m} =    
    \frac{2  Z \Lambda_{QCD}^6}{\Delta m M_*^{4+N} V_N},
    \label{decayrate}
\end{equation}
where we plugged-in the parameter $\alpha$ from Eq. \eqref{alpha}.
Using the expression (\ref{Puseful}), it is also useful to present the rate as
\begin{equation}
    \lambda_n \,  \sim  \,  m_n  \left (\frac{\alpha^2}{m_n^2} \right )  (m_nR)^{N_R} \,.
    \label{nrate1}
\end{equation}
Now, we demand that $\tau_n = 1/\lambda_n > 10^{30}$ y.
This leads to the following constraint, 
 \begin{equation}
    M_*^{4+N}> 10^{21} ({\rm GeV} \: R)^{N_R} M_f^{2+N} {\rm GeV}^2 \,,
    \label{starbound}
\end{equation}
where we have expressed $V_N$ through $M_P \sim 10^{19}$GeV and $M_f$ and  used the fact that $m_n \simeq $ GeV and $ \Lambda_{QCD} \simeq 0.3$ GeV.
We also used the neutron lifetime bound (\ref{taun}) in GeV units. 
  
We now wish to apply this bound to different cases distinguished by the number and size of relevant dimensions.
For this, let us first make the concept of relevance more quantitative.   
We have $N_R$ relevant dimensions of size $R$ and $N-N_R$ subdominant ones of radii $\tilde{R}$. 
Both of these categories contribute to (\ref{masterequation}). 
The distinction is that the contribution of the subdominant dimensions to the rate (\ref{nrate1}) is less.
That is, 
   \begin{equation}  \label{RtildeR} 
  (m_n\tilde{R})^{N-N_R} \,  \ll \,    (m_nR)^{N_R} \,. 
\end{equation} 
Then taking into account the relation (\ref{masterequation}) and the fact that $V_N = (R^{N_R}\tilde{R}^{N-N_R})$, we get
     \begin{equation}  \label{COND} 
    (m_n\tilde{R})^{N-N_R}  \ll  \frac{M_P}{M_f} \left (\frac{m_n}{M_f} \right )^{N /2}.
\end{equation}

We can now apply the above to the following cases. 
The first is the case $N_R =1$ with just one relevant extra dimension of radius $R$.
The radius $R$ and $M_f$ are free parameters.
Therefore, we set both to their current experimental bounds, given by equations (\ref{RRR}) and (\ref{LHC}). 
The results for different total numbers $N$ of extra dimensions are shown in Table \ref{table:MstaronedominantR}.

\begin{table}[ht!]
\centering
\begin{tabular}{|x{2.5cm}|x{2.5cm}|}
\hline
 $N$ & $M_*[{\rm GeV}]$\\
 \hline
 3  & $>3 \cdot 10^7$ \\
 4  & $> 1 \cdot 10^7$ \\
 5  & $>5 \cdot 10^6$ \\
 6  & $>3 \cdot 10^6$ \\
\hline
\end{tabular}
\caption{Bound on $M_*$ for one dominant $R$ with $M_f = 10$ TeV and $R = 30 \mu$m}
\label{table:MstaronedominantR}
\end{table}

Notice that condition (\ref{COND}) implies that for $N_R =1$ and the above choice of $R$ and $M_f$, we must have $N> 2$. 
In other words, with only two extra dimensions and $M_f \sim 10$TeV,  it is not possible to have one much shorter than
the other without conflicting with observations. 

The second example we study is the case with $N$ equal size extra dimensions. 
That is, in this case, $N_R \, = \, N$. Correspondingly, using (\ref{masterequation}), 
the r.h.s. of \eqref{starbound} can be expressed via  $M_P \sim 10^{19}$GeV, and the bound acquires a simple numerical form, 
\begin{equation}
    M_*^{4+N} \, > \, 10^{59} {\rm GeV}^{4+N} \,. 
    \label{starboundN}
\end{equation}

The resulting bounds on $M_*$ and corresponding values of $R$ are listed in Table \ref{table:Mstarequalsize}. \\

\begin{table}[ht]
\centering
\begin{tabularx}{0.45\textwidth} { 
  | >{\centering\arraybackslash}X 
  | >{\centering\arraybackslash}X 
  | >{\centering\arraybackslash}X 
  | >{\centering\arraybackslash}X 
  | >{\centering\arraybackslash}X | }
 \hline
 $N$ & $R[\mu {\rm m}]$ & $M_*[{\rm GeV}]$\\
 \hline
 2  &  $1.1$                & $>7 \cdot 10^9$\\
 3  &  $1.6 \cdot 10^{-5}$  & $>3 \cdot 10^8$ \\
 4  &  $5.5 \cdot 10^{-8}$  & $>2 \cdot 10^7$ \\
 5  &  $2 \cdot 10^{-9}$    & $>4 \cdot 10^6$ \\
 6  &  $2.2 \cdot 10^{-10}$ & $>8 \cdot 10^5$ \\
\hline
\end{tabularx}
\caption{Bound on $M_*$ for equal size extra dimensions.}
\label{table:Mstarequalsize}
\end{table}

Another instructive choice is $M_* \sim M_f$.  In this case, (\ref{starbound}) becomes
\begin{equation}
    M_f^2 \, > \,  10^{21} ({\rm GeV} R)^{N_R} {\rm GeV}^2 \,.
    \label{Fbound}
\end{equation}
Since $N_R \leqslant N$,  from  (\ref{masterequation}) we have  $M_f \leqslant (M_P^2/(2\pi R)^{N_R})^{1/(N_R +2)}$.
Together with this inequality, equation  (\ref{Fbound}) becomes a bound on $R$. 
For example, for $N_R=1$ we get
\begin{equation}
    R \,\lesssim 10^2 {\rm GeV} ^{-1}  \sim 10^{-8} \mu {\rm m} \,.
\end{equation}
Equivalently, the bound on $M_f$ is
\begin{equation}
    M_f \, \gtrsim \, 10^{12} {\rm GeV} \,.
\end{equation}

\section{Proton Decay}

The possibility of neutron oscillations into a bulk fermion opens the door to proton decay. 
If the high-dimensional mass of $\Psi$ is less than the mass difference between the proton and the sum of electron and neutrino masses, the tree-level decay process is possible. Via a virtual neutron exchange, the proton can decay into a positron, the SM neutrino, and $\Psi$,  
\begin{equation} 
    p \rightarrow e^{+} +  \nu + \Psi.
\end{equation} 

\begin{figure}[!ht]
    \centering
    \includegraphics[width=\columnwidth]{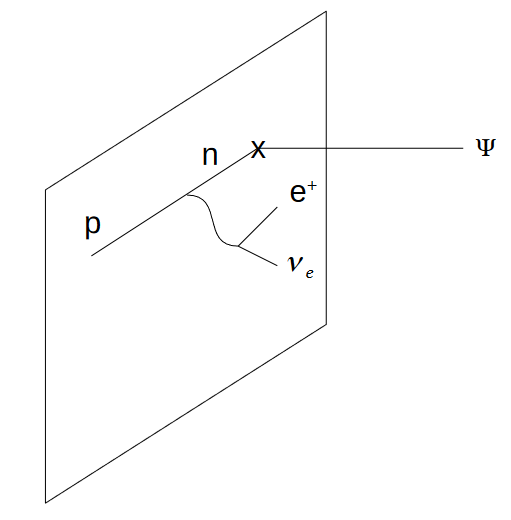}
    \caption{Proton Decay via virtual neutron exchange. The SM particles are confined to the brane, whereas the extra-dimensional particle, $\Psi$, propagates in the bulk.}
    \label{ProtonDecay}
\end{figure}
The process is depicted in Fig. \ref{ProtonDecay}.
The rate of the process is given by
\begin{equation} 
    \lambda_p \, \sim  \,   m_p \, \left (\frac{m_n}{v} \right )^{4} \, \left (\frac{\alpha}{m_n} \right )^{2} \, (m_p R)^{N_R}.
\end{equation} 
It is useful to compare this rate to the neutron disappearance rate (\ref{nrate1}). 
Taking into account that $m_n \simeq m_p$,  we obtain that the proton decay rate is suppressed relative to the neutron rate by an additional factor  $ \left (\frac{m_n}{v} \right )^{4} \sim 10^{-9}$ coming from the $W$-boson exchange: 
\begin{equation} 
    \lambda_p \, \sim  \,  \lambda_n  \left (\frac{m_n}{v} \right )^{4}\, \sim \lambda_n 10^{-9} \, .
\end{equation} 
At the same time, the current bound on the proton lifetime for this channel, $\tau_p > 10^{30}$y~\cite{Learned:1979gp}, is the same as for the bound on neutron disappearance (\ref{taun}). 
Correspondingly, the bound on the proton lifetime is automatically satisfied, as long as the neutron bound is fulfilled.  
  
Because of this, in terms of $M_*$  the proton gives a milder bound.  
E.g., assuming $M_f \sim 10$TeV,  the bound is
\begin{equation} 
    M_*^{4+N} >  10^{12} (m_p R)^{N_R} \, M_f^{2+N} {\rm GeV}^2\,.
\end{equation} 
Therefore, the improved accuracy of neutron disappearance experiments would appear 
more promising for testing the discussed scenario. 

 
 \section{Kaluza-Klein Spectroscopy from Free Neutron Oscillations}

We now wish to discuss another experimentally-motivated domain of our framework. 
In this domain, the dominant effect is the oscillation of a free neutron into hidden dimensions. 
In this regime,  the bulk partner of the neutron,  $\Psi$, has a high-dimensional mass, $\mu$.
If the mass is above the energy levels of the nuclear neutron, the stability of nuclei is unaffected.
Similarly, if $\mu$ is above the proton mass, there is no bound from proton decay.
Thus, if $\mu$ is within a window, $\sim $ MeV below the mass of a free neutron,
only the free neutron is experiencing oscillations into extra dimensions. 
The transition amplitude can correspondingly be much larger. 

Notice that for dimensions that satisfy $1/R \, \ll \, |m_n-\mu|$, the number and the level-splitting of the available  KK modes to which the neutron can oscillate is essentially the same as for $\mu \, = \, 0$. 

For $N_R = 2$ this level-splitting is sufficiently small to be scanned by an external magnetic field available in current laboratory setups.
This allows for potentially-observable resonant $n-\Psi$ oscillations
in correlation with the features of the KK spectrum. 
Let us discuss this effect. 

The amplitude of the oscillation of the neutron into the bulk particle is suppressed by the minimal mass splitting between the neutron and the nearest KK mode $\Delta m$.
At the same time, the amplitude is enhanced by the  KK multiplicity factor $Z$.  
  
The oscillation of a free neutron is effectively mapped onto a $2\times 2$ problem, in which the neutron mixes with a single state with the nearest mass,
\begin{equation}
    \begin{pmatrix}
        m_n & \sqrt{Z}\alpha  \\
        \sqrt{Z} \alpha & m_n + \Delta m 
    \end{pmatrix}.
\end{equation}
   
In the absence of a parameter that could scan the neutron energy with a precision of $\Delta m$, the oscillation amplitude is suppressed as given by (\ref{Puseful}).   
However, an external magnetic field can serve as a scanner.  
A neutron placed in a magnetic field $\vec{B}$, because of its magnetic moment $\vec{\mu}_n$, experiences an additional shift in the energy level equal to $\epsilon \, = \, \vec{\mu}_n\vec{B}$.
  Correspondingly, the mixing matrix becomes
\begin{equation}
    \begin{pmatrix}
        m_n  + \epsilon& \sqrt{Z}\alpha  \\
        \sqrt{Z} \alpha & m_n + \Delta m   
    \end{pmatrix}.
\end{equation}
Notice that, regardless of the value of the magnetic field, in the above mixing matrix we must always assume  $\epsilon  \lesssim  \Delta m$.
If we increase the magnetic field in the way that $\epsilon  >  \Delta m$, the whole problem will be shifted to the  KK level which is closest to the shifted energy of the neutron $m_n + \epsilon$.
Thus, an increase in the  magnetic field above the resonance value, 
\begin{equation} \label{Br}
    B_r \equiv \Delta m/\mu_n, 
\end{equation} 
results in the effective shift  $m_n  \rightarrow m_n \,  + \,  \epsilon$. 

Because of this, for the array of magnetic field values, the neutron energy comes in resonance with some of the KK levels, and an enhanced oscillation amplitude is achieved. 

For $N_R\, =\, 2$ the levels are not exactly equally spaced. Because of this,  the effect is not strictly periodic but has a clear repetitive pattern
fully reflecting the KK spectrum.

This feature is very different from the theories in which the neutron has a single oscillation partner, e.g., such as the dark neutrons from the hidden standard model copies.
We shall make a more explicit comparison with such scenarios later.

For currently available experimental setups, $\epsilon$ is a very small quantity.
For example, for earth's magnetic field, $|\vec{B}_e| \simeq 0.5$G, we have $\epsilon_e \sim 10^{-12}$eV.
With the presently available most powerful artificial magnets, it can be increased by a few orders of magnitude.

This is sufficient for scanning the KK spectrum for $N_R =2$ in an interesting range 
of radii.   For example, for  the largest possible value of extra dimensions
$R = R_{max}$, given by experimental bound (\ref{RRR}), the splitting of closely spaced KK modes is $\Delta m  \equiv  1/(2m_nR_{max}^2) \simeq   2 \times  10^{-14}$ eV. 
This value falls within the territory testable by currently available experimental setups. 
 
For  $\alpha < |\epsilon - \Delta m| < \Delta m $, the oscillation amplitude is
\begin{equation} \label{AMP}
    A \, \simeq \, \frac{\alpha^2}{|\epsilon - \Delta m|^2} \,,  
\end{equation} 
and the oscillation frequency is given by $\omega = |\epsilon - \Delta m|$.
For the resonance regime,  $\alpha \leq  |\epsilon - \Delta m| $, the amplitude becomes order one, and the frequency is set by $\omega \sim \alpha$. 
   
Notice again that for the regime $\epsilon \gg  \Delta m$,  we are shifting to a new  KK level which is closest to $m_n + \epsilon$, and the excess of $\epsilon$ gets effectively reabsorbed in the shift of $m_n$.
 
We shall now confront the above dynamics with some experimental data. 
In order to establish a dictionary,  we shall recast the expression for the amplitude (\ref{AMP}) into a more user-friendly form.
For this, we take into account that 
$\epsilon \simeq 6 \times 10^{-14} B \frac{{\rm eV}}{\mu {\rm T}}$~\cite{Tiesinga:2021myr} and
that for $R =R_{max}$, we have $\Delta m \simeq 2 \times  10^{-14}$ eV. 
We can then write
\begin{equation} \label{AAAN}
    A \sim   \frac{10^{27} \,\alpha^2 \, {\rm eV}^{-2}}{|3 \frac{B}{\mu {\rm T}}  - \frac{R_{max}^2}{R^2}|^2}   \, .
\end{equation}  
Since for resonant values of $B$, the defining factor is the
difference of the two terms in the denominator, we pay 
more precise  attention to them,  while examining the 
order-one overall numerical coefficients less closely.     
We shall use this formula for interpreting the two experimental results. 

The first one~\cite{nEDM:2020ekj}, the ultra-cold neutron storage experiment, constrains the disappearance of a neutron for two separate values of the magnetic field: 
$B \simeq 10.20 \pm 0.02 \mu$T  and $B \simeq 20.39 \pm 0.04 \mu$T. 
   
If we assume $R = R_{\rm max}$, then for the above 
values of the magnetic field, the first entry in the denominator (\ref{AAAN}) is larger than the second one by 
factors of $30$ and $60$, respectively. 
This implies that the system will shift to a higher KK level, absorbing the 
extra contribution from the magnetic energy into the mass of a new KK partner. 
Excluding miraculous coincidences of finer cancellation,  
the optimal KK level will satisfy  
\begin{equation} \label{OPTIMAL} 
 |\epsilon - \Delta m|
\sim \Delta m \,.
\end{equation}

Notice that the error in $\delta \epsilon = \epsilon \delta B/B$, 
due to an inaccuracy of the magnetic field, which in both cases is $ |\delta B/ B| \simeq 2\times 10^{-3}$, 
is not  sufficient for further reducing the difference
$|\epsilon - \Delta m|$ significantly. 
  
 Now, it is clear that the relation (\ref{OPTIMAL}) will persist for 
 $R \, < \, R_{\rm max}$. Indeed, for 
 $\frac{R_{max}^2}{R^2}   \ll 3 \frac{B}{\mu {\rm T}}$ 
 the level will always get shifted to the one that satisfies 
 (\ref{OPTIMAL}).  At the same time, for $\frac{R_{max}^2}{R^2} \, >  \, 3 \frac{B}{\mu {\rm T}}$, the same condition is  satisfied without any shift.  Correspondingly,  the relation (\ref{OPTIMAL}) is satisfied 
 for the entire parameter space probed by the 
 experiment of~\cite{nEDM:2020ekj}. 
 Thus, for fitting the data,  the equation (\ref{AAAN}) can 
 be approximated as, 
 \begin{equation} \label{AAAFIT}
    A \sim   10^{27} \,\alpha^2 \, \frac{R^4}{R_{max}^4} \, {\rm eV}^{-2} \, .
\end{equation}

Putting everything together, we get the following constraint on our parameters, 
\begin{equation} \label{ACON}
    \, \alpha \,  \frac{R^2}{R_{max}^2}  \lesssim \,   10^{- 16} {\rm eV} \,.
\end{equation}
Expressing $\alpha$ in terms of $M_*$ and $M_f$, we can write, 
\begin{equation} \label{ACONM}
    \, \frac{10 {\rm TeV}}{M_f} \left (\frac{M_f}{M_*} \right )^{2+N/2}  \,  \frac{R^2}{R_{max}^2}  \lesssim \,  \frac{1}{3} \,.
\end{equation}

  Interestingly, by taking $M_f$ at its current experimental bound
 (\ref{LHC}) of $M_f \sim 10$TeV, while at the same time taking $M_*$ at its theoretical one, $M_* \sim M_f$, 
 equation (\ref{ACONM}) translates as the current experimental bound on $R$ (\ref{RRR}) imposed by the checks of 
 Newtonian gravity.  Having a rigorous bound on the extra-dimensional setup from the cold neutron experiments is rather remarkable.

We shall now move to the second experiment~\cite{Ban:2023cja}, with an ultra-cold neutron beam. 
The important insight from this measurement is that the authors scanned a wide range of the magnetic field
within the interval in between $B_{min} \simeq 50 \, \mu$T and $B_{max} \simeq 1100 \,  \mu$T,  with a step of $\Delta B = 3\mu$T. 
In order to translate the results of this experiment into bounds on our parameters, we must distinguish the following cases.

In the regime
\begin{equation} \label{BMAX}
\frac{R_{max}^2}{R^2} \,  \gg \,  3 \frac{|B_{max} - B_{min}|}
{\mu {\rm T}}  \simeq 3150 \,, 
\end{equation}
 or equivalently  
$\Delta m \gg \epsilon_{max} - \epsilon_{min}$, 
our transition amplitude is essentially the same as without 
the magnetic field,  (\ref{AAAFIT}), and the bound is
\begin{equation} \label{ABB}
    \, \alpha \,  \frac{R^2}{R_{max}^2}  \lesssim \,   10^{- 15} {\rm eV} \,.
\end{equation}
It has the same form as (\ref{ACON}) but 
subject to (\ref{BMAX}). 

In the opposite case,
\begin{equation} \label{BMAX1}
\frac{R_{max}^2}{R^2} \,  <  \,  3 \frac{|B_{max} - B_{min}|}
{\mu {\rm T}} \,,
\end{equation}
 the bound depends on the 
ratio between $\Delta m$ and the scanning step 
$\Delta \epsilon$.  
 In the denominator of expression (\ref{AMP}) the smallest 
  of the two will enter. 
Because of this, for  $R_{max}^2/R^2 <   9$, the  bound is again 
  given by  the expression (\ref{ABB}). 
  
   For $R_{max}^2/R^2  >  9$, the situation is different 
  since the   $\Delta m$ term in (\ref{AMP})  
  can be compensated by the accuracy of $\Delta \epsilon$.

 The magnetic term  $\epsilon$ can be gradually cranked up in small increments $\Delta \epsilon$ all the way to $\Delta m$, sooner or later the resonant transition will take place with $|\epsilon - \Delta m| \simeq \Delta \epsilon$.  
In this case, the absolute value of $\Delta m$ (and thus dependence on $R$) drops out from the transition amplitude and we have 
\begin{equation} \label{AMPE}
       A \, \sim  \, \frac{\alpha^2}{|\Delta \epsilon|^2} 
  \sim  \alpha^2 10^{25} \, {\rm eV}^{-2} \, .    
\end{equation} 
 This imposes the following $R$-independent constraint on  $\alpha$
 \begin{equation} \label{Ascan}
    \, \alpha \,  \lesssim \,  10^{- 14} {\rm eV} .
\end{equation}

\begin{figure}[!ht]
    \centering
    \includegraphics[width=\columnwidth]{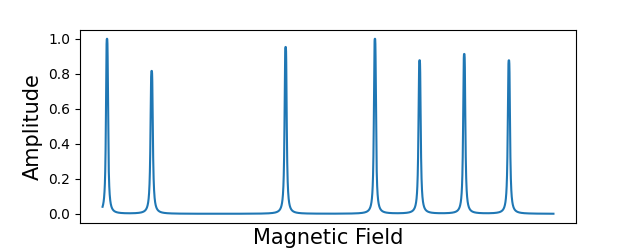}
    \caption{A qualitative sketch of the transition amplitude as a function of the magnetic field. 
  The resonance transitions take place with 
  steps $\Delta B = \Delta m/\mu_n$.  The differences in 
  the heights are due to different degeneracy factors of different 
  KK levels.  
   The repetitive but not strictly periodic behavior 
   is due to the nonuniform population of KK levels. 
    }
    \label{fig:FreeNeutronResonance}
\end{figure}

\begin{figure}[!ht]
    \centering
    \includegraphics[width=0.8\columnwidth]{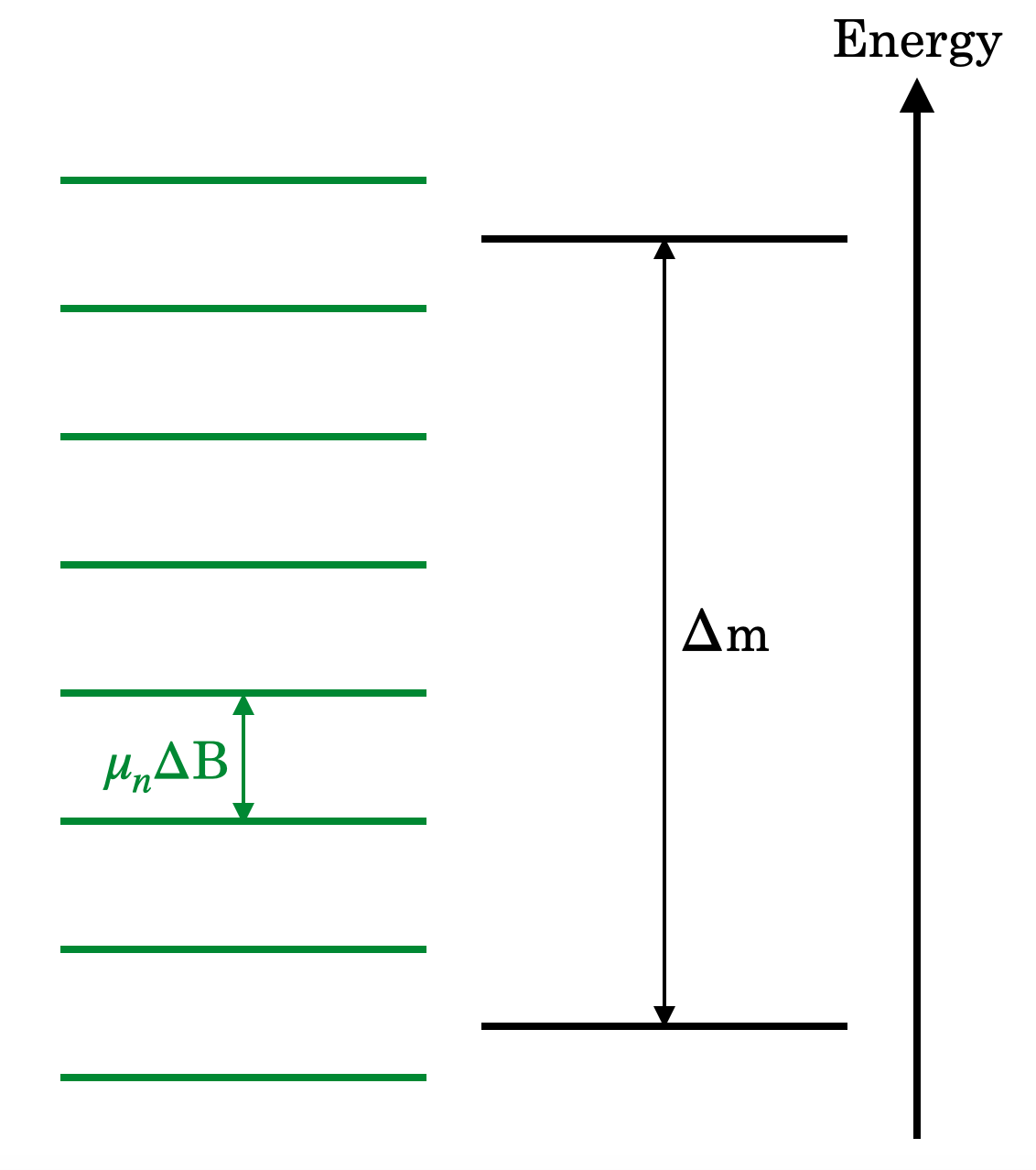}
    \caption{Scanning with the magnetic field between two KK-levels. A resonance occurs when the shifted energy level of the neutron comes close to a KK state with a precision of $\mu_n \Delta B$.
    }
    \label{fig:ResonanceLevels}
\end{figure}

In summary, the characteristic signature of neutron oscillations into extra dimensions, which makes it very different from other proposals, is the recurrence of the resonance amplitude for multiple values of the magnetic field with steps $\Delta B \,  = \, \frac{\Delta m}{\mu_n}$, where $\Delta m$ is the mass splitting between the nearest KK levels.  
Such recurrent resonance transitions are the prediction of the 
extra dimensional scenario, see Fig. \ref{fig:FreeNeutronResonance} and Fig. \ref{fig:ResonanceLevels}.  

The experiment~\cite{Ban:2023cja} is effectively probing the interval
\begin{equation}
    0.8 \mu\textrm{m} < R < 10\mu \textrm{m}\, 
\end{equation}
and correspondingly impose the bound (\ref{Ascan}) on $\alpha$.

 Extensions of the upper and lower bounds of this interval can be achieved by an increase of the magnetic field-range and the decrease of the step size, respectively.  For example, a decrease of the step size 
 to  $1/3 \mu$T  would bring the upper bound to the
 level of (\ref{RRR}) imposed by the measurements of Newton's law. 
 An increase of the range of the scanned magnetic field would correspondingly allow 
 to probe smaller sizes of extra dimensions. 
 
The above analysis makes the general tendency clear. 
The experiments with finer scanning of wider ranges of the magnetic field can provide deeper probes of physics of extra dimensions.  Strikingly, the bounds (\ref{ACON}), (\ref{ABB}) and 
(\ref{Ascan}) indicate that these experiments already probe the domain 
motivated by the Hierarchy Problem.

\section{Comparing with oscillations into hidden copies of the neutron}    
 
In this section, we wish to confront the presented scenario with the previous proposal~\cite{Dvali:2009ne} in which the neutron also mixes with multiple partners  $n_i,~i=1,2, ..., N$ from $N$ hidden sectors
($N$ not to be confused with the same notation of number of dimensions in ADD).
These partners represent neutron-like particles belonging to $N$ copies of the SM. 
The copies are related by an exact permutation symmetry.
Similarly to the ADD model, such a scenario was originally motivated by the Hierarchy Problem, since the existence of many SM copies lowers the cutoff of the theory~\cite{Dvali:2007hz,Dvali:2007wp}. 
   
In a sense, this scenario is a ``Fourier transform" of the ADD solution meaning that the dilution of gravity takes place in the space of species.
Correspondingly, solutions to certain puzzles offered by ADD, also find counterparts in this framework. 
For example, a ``dilution"  of the neutrino mass in the space of species can be achieved 
\cite{Dvali:2009ne, Ettengruber:2022pxf}. Also, the particles in the hidden copies of SM can be   
dark matter \cite{Dvali:2009fw, Dvali:2009ne} \footnote{In fact, this idea was originally proposed within the ADD framework, with the role of exact copies of SM played by the 
parallel folds of our brane \cite{Arkani-Hamed:1999rvc}}.

As discussed in~\cite{Dvali:2009ne}, one of the phenomenological consequences of the
scenario with $N$ SM copies can be the oscillation of a neutron into the hidden partners, since the mixing $\alpha \sum_{i\neq j}\bar{n}_in_j$ is permitted by the gauge symmetries of all sectors. 
The only constraint,  $\alpha  < m_n/N$,  comes from unitarity.  

This scenario exhibits some crucial differences from the present case. 
Because of the exact permutation symmetry, the mass matrix has the form, 
\begin{equation}
    \begin{pmatrix}
        1 & 0 & 0 & 0\\
        0 & 1 & 0 & 0\\
        0 & 0 & 1 & 0\\
        0& 0 & 0 & 1
    \end{pmatrix}  (m_n - \alpha) \, 
    +  
   \begin{pmatrix}
        1 &  1 & 1 & 1\\
        1 & 1 & 1 & 1\\
        1 & 1 & 1 & 1\\
        1 & 1 & 1 & 1
    \end{pmatrix} \, \alpha\,.    
\end{equation}   
Due to this structure, the oscillation dynamics can be reduced to a $2\times 2$ problem in which the neutron from our copy, $n_1$, mixes with the state $n_h \equiv \frac{1}{\sqrt{N-1 }} \sum_{j\neq 1} n_j$, via the following mass matrix, 
\begin{equation}
    \begin{pmatrix}
        m_n & \alpha \sqrt{N-1}\\
          \alpha \sqrt{N-1} & m_n + \alpha \sqrt{N-2}\\
    \end{pmatrix}  \,.  
\end{equation}   
The rest of the orthogonal $N-2$ states decouple. 

The resulting disappearance probability is
\begin{equation} 
    P(t) \simeq \frac{4}{N} \sin^2\left (\frac{N \alpha t}{2} \right )\,.
\end{equation} 
This picture is very different from the case of a neutron mixing with the KK tower. 
The main difference is that the neutron possesses a single oscillation partner in the form of a state $n_h$.  
For neutrons in all possible energy states, this partner is unique and is fixed by the theory.

The phenomenology of free neutron oscillations is correspondingly very different
from the KK case, in particular, in the magnetic field dependence of the resonant amplitude. 
In case of an oscillation into a single hidden partner, the amplitude peaks around the resonant value of the magnetic field $B_r \, = \, \alpha \sqrt{N-2}/\mu_n$, and diminishes in both directions as a function of $|B - B_r|$. 
   
This behavior goes in sharp contrast with the present scenario in which a neutron oscillates into the KK tower.  
As already discussed, in this case, a neutron of arbitrary energy $m_n + \epsilon$ finds its oscillation partner among the KK states that are closest to it. 
Because of this effect, the amplitude is recurrent in $B$ with the steps  $\Delta m/\mu_n$. 
     
This difference applies to a picture with an arbitrary number of SM copies including the special case of $N=2$. 
This case is usually referred to as the mirror SM.
The oscillation between the neutrons of the two copies was studied earlier in~\cite{Berezhiani:2005hv}.
The additional difficulty that arises in this case is the exact degeneracy of the diagonal masses between $n$ and its mirror partner $n'$.
Because of this, the oscillations are suppressed by potentials arising in our sector due to the environment, such as the earth's magnetic field which gives a large level splitting. 
Therefore, the transition can take place only if one assumes the presence of an analogous magnetic field in the hidden sector, which is outside of any experimental or theoretical control, as it fully depends on the nature of the hidden sector.   
Regardless of this issue, assuming the proper conditions, the resonance can only occur for one specific value of the magnetic field which happens to match the hidden one.  
This is different from the extra-dimensional oscillations of neutrons for which the resonances happen for discrete values of the magnetic field that are synchronized with the KK spectrum.

\section{Neutron lifetime measurements}

The possibility of free neutron oscillations into extra dimensions also creates an urgency for more precise measurements of its lifetime. 
In fact, some authors have argued that already existing measurements may indicate a discrepancy that supports the existence of new channels of neutron disappearance.  
  
On the one hand,  we possess the data coming from so-called ultra-cold neutron lifetime measurements.
They account for the missing neutrons from a given number of free neutrons.
The reported lifetime of the neutron is $\tau_1 = 878s$~\cite{UCNt:2021pcg,Ezhov:2014tna,ParticleDataGroup:2022pth}.
On the other hand, we have the data from beam experiments that look for the protons resulting from ordinary decay of neutrons caused by the weak interaction.
These experiments report a lifetime of $\tau_2 = 888s$~\cite{Yue:2013qrc,Nico:2004ie,ParticleDataGroup:2022pth}. 

This difference between the reported neutron lifetimes
 prompted some attempts to address it in the context of oscillations to mirror neutrons~\cite{Berezhiani:2005hv,Berezhiani:2009ldq,Berezhiani:2018eds,Tan:2019mrj}, or decays of neutrons into other hypothetical particles~\cite{Fornal:2018eol} (for some cosmological implications, see, \cite{Karananas:2018goc}).
However, doubts about the rigor of the discrepancy have also been raised~\cite{Czarnecki:2018okw}.

Scrutinizing the validity of the puzzle is beyond the goal of the present paper. 
Regardless, the neutron oscillations into extra dimensions provide an additional motivation for improved precision measurements of the neutron lifetime.

 \section{Baryon and Lepton Numbers} 

Notice that the mixing of the SM neutrino with a bulk fermion
(\ref{eq:neutrinodiracExtraDim1}) preserves lepton number symmetry since $\Psi$ can be assigned one unit of lepton number. 
Likewise, the mixing of the neutron with the bulk fermion 
$\Psi$ (\ref{actionNPsi}), fixes the baryonic charge of $\Psi$ as equal to one. 
In the case where neutrino and neutron mix with the same $\Psi$, one combination of the baryon and lepton number is preserved depending on whether both mix with $\Psi$ or
with relative conjugates.
Mixing with the same $\Psi$ leaves $B+L$ symmetry unbroken and breaks $B-L$.
Mixing with the conjugates $\Psi$ and $\Psi^*$ has the opposite effect.  
Notice, as it was discussed in~\cite{Arkani-Hamed:1998sfv}, the $B-L$ symmetry can be gauged in the bulk, resulting in gravity competing forces from the exchange of 
a $B-L$ gauge boson. 
If $B-L$ is gauged, one of the mixing operators must be suppressed by the vacuum expectation value 
of the field that Higgses it. 
   
We would also like to comment on the case when the bulk particle $\Psi$ that mixes 
with neutron possesses a Majorana mass of the form $\mu_{M} \Psi C \Psi$.
This term breaks the baryon number by two units.
Due to this, the exchange of $\Psi$ can result in neutron-anti-neutron oscillations.
In this case, the neutron disappearance rate becomes correlated with the $n-\bar{n}$ oscillation.  This case will not be discussed here further.

\section{Cosmology} 
 
We would like to briefly comment on some straightforward  
cosmological implications and constraints on the present scenario. 
 
In the large extra-dimensional theory, there exists a standard list of constraints shared by the bulk species~\cite{Arkani-Hamed:1998sfv}. 
In particular, an important 
bound is coming from Big Bang Nucleosynthesis (BBN). 
In this consideration, the key control parameter is the ``normalcy temperature". The point is that at BBN temperatures the production of bulk species must be sufficiently suppressed in order not to 
interfere with the standard evolution of the Universe. 
The bulk species must neither enter in thermal 
equilibrium with the SM particles nor be overproduced.    
  
In the case of a massless bulk fermion $\Psi$
mixing with the neutrino, such constraints were discussed 
in~\cite{Dvali:1999gf}.  The bound in the present case is milder 
 since the neutron mixes with the sector of the KK 
 tower which is much heavier than the BBN temperature. 
 
  In particular, in the parameter regime relevant 
  for free neutron oscillations into the bulk fermion, the thermal production of 
 $\Psi$ due to the rescattering of quarks via the four-fermi operator  (\ref{actionNPsi}) is effectively shut off below the temperature of its mass $\mu \sim$ GeV. 
 
    Besides the constraints, cosmology can provide some strong motivations for the existence of a neutron portal into extra dimensions 
    in the form of its bulk partner $\Psi$. 
     The first one is baryogenesis via the mechanism introduced in~\cite{Dvali:1999gf}. The idea is that the excess of baryonic change in our sector (i.e., particles inhabiting the SM brane) is generated by its loss into extra dimensions. 
     
    Even though a transition of a neutron into an extra 
    dimension does not violate baryon number, baryogenesis
    can still take place, provided the other two  Sakharov's 
    conditions ($CP$-violation and an out-of-equilibrium state) 
    are satisfied. This is because the bulk species  can      
    transport baryon number away from our SM sector and ``hide" 
 it in the KK states. 
 
 In our case, such a process 
 can be realized as an out-of-equilibrium 
 conversion of our baryons into the fermions $\Psi$.  Since $\Psi$ is extremely weakly interacting, the inverse decays are highly suppressed.  
The overall effect can be a generation of 
 net baryon number in our sector. 
  Of course, the exact amount of the missing baryonic charge 
  is carried by the bulk species. 
  However, this charge is effectively inaccessible 
  for our measurements. Hence, the Universe appears to 
  be asymmetric in baryon number.

   The second natural implication of the bulk particle $\Psi$ is that 
  its KK states can play the role of dark matter. 
  The generic feature of bulk modes is that they interact 
  with SM species with gravitational strength and, correspondingly, are extremely long-lived. This makes them viable candidates 
  for dark matter.    
   The possibility that KK gravitons of large extra dimensions 
   can be dark matter, was originally proposed in~\cite{Arkani-Hamed:1998sfv}  (a more recent discussion 
   of this idea can be found in \cite{Gonzalo:2022jac}). 
  
   Similarly, in our scenario, $\Psi$ can be a dark matter candidate. 
   Of course, the construction of a fully viable and predictive scenario  
   requires a more detailed cosmological investigation 
   of the parameter space of our setup. This 
   is beyond the goals of the present paper.

\section{Conclusions and outlook}

In this paper, we pointed out that neutron experiments 
can provide a unique spectroscopy of extra dimensions. 
 
Neutrons can play a crucial role in exploring new 
particle species that interact with the SM
ultra-feebly: with a strength comparable to that of gravity.   
Especially motivated are the frameworks in which the number of ultra-feebly interacting species 
is large.  This is because such frameworks address the Hierarchy Problem 
by lowering the cutoff of the theory according to (\ref{species}). 
The two extreme representatives of such theories are the ADD model 
of large extra dimensions~\cite{Arkani-Hamed:1998jmv,Arkani-Hamed:1998sfv}
and the framework with multiple exact copies of the SM~\cite{Dvali:2007hz,Dvali:2007wp,Dvali:2009ne}. 
In the latter framework, oscillations of a neutron into hidden copies of the SM have been studied in~\cite{Dvali:2009ne}.  
However, oscillations of a neutron into large extra dimensions have not been studied previously. In the present paper, we attempted to fill this gap.    
  
Within the ADD framework, we studied the oscillations of a neutron into a fermion  $\Psi$ propagating 
in large extra dimensions.  In particular, $\Psi$  can be the same bulk fermion 
that endows the SM neutrino with a naturally small mass via the 
mechanism of~\cite{Arkani-Hamed:1998wuz,Dvali:1999cn}. 
The consequence of this mechanism is an oscillation of the neutrino into the KK  tower of the bulk sterile fermion~\cite{Dvali:1999gf}.

  In the present setup, a similar oscillation into the KK species of bulk fermion 
  is experienced by the neutron. 
  However, neutron oscillations exhibit certain features that make 
  them subject to special  interest  for a wide range of experiments and phenomenological 
  constraints.  
   A unique feature of a neutron mixing with a bulk species  
    is that because of the high density and degeneracy of the
  KK spectrum, neutrons in various energy states, both free or within nuclei, 
  find their oscillation partners.  The threshold is set by the mass of the bulk fermion $\Psi$ which is a theory parameter. 

For the mass below the energy of a nuclear neutron,  the main 
effect is neutron disappearance from nuclei resulting in its de-excitation  
and the emission of a hard photon.  This imposes a variety of bounds 
on the parameters of the theory.   The associated proton decay rate 
is suppressed by another nine orders of magnitude and is a subdominant effect. 
 Therefore, the avenue for tightening the constraints is via improving the precision 
 of neutron disappearance experiments as well as in scanning a larger diversity of samples
 since the resonant energy levels depend on the features of the Kaluza-Klein spectrum. 
 
    For the mass of the bulk partner in a window between the energies 
    of a bound and a free neutron, 
   the nuclei are stable. However, the resonant transition can be observed for a free neutron.
    This gives an exciting possibility of mapping the Kaluza-Klein spectrum 
  by using a magnetic field as a scanner.  Unlike the scenarios in which the neutron has 
  a single oscillation partner,  in the extra-dimensional case, 
  the resonance takes place recurrently as a function of a magnetic field. 
   That is, the resonant values of the magnetic field are quantized in one-to-one 
   correspondence with the KK spectrum.  
  
   We analyzed our findings in the light of two recent experiments:
    the  ultra-cold neutron storage experiment~\cite{nEDM:2020ekj} and the experiment with an ultra-cold neutron beam~\cite{Ban:2023cja}. 
   We found that both experiments impose bounds on the parameters of extra dimensions. 
In fact, the bounds, such as (\ref{ACON}), (\ref{ABB}) and 
(\ref{Ascan}), are already within the domain motivated by the Hierarchy Problem.

Future experiments with improved accuracy and a wider range of the scanned
    magnetic field  will give a unique possibility of performing precision 
    Kaluza-Klein spectroscopy in neutron labs.  \\
  
 {\textsl{\bf Acknowledgments}}\;---\; This work was supported in part by the Humboldt Foundation under the Humboldt Professorship Award, 
  by the European Research Council Gravities
Horizon Grant AO number: 850 173-6,
 by the Deutsche Forschungsgemeinschaft (DFG, German Research Foundation) under Germany's Excellence Strategy - EXC-2111 - 390814868, Germany's Excellence Strategy under Excellence Cluster Origins and the Sonderforschungsbereich SFB1258.\\

 Disclaimer: Funded by the European Union. Views and
opinions expressed are however those of the authors only
and do not necessarily reflect those of the European Union
or European Research Council. Neither the European
Union nor the granting authority can be held responsible
for them.

\section{Appendix}

\subsection{Mass Splitting in KK Tower for Equal Size Extra Dimensions}
We show that for $N \geq 2$ with all radii equal to $R$, the mass splitting of the KK tower (around the neutron mass) is proportional to
\begin{equation}
    \delta m \sim \frac{1}{R^2 m_n}.
\end{equation}
This is correct for both the KK tower originating from a massless or a massive extra-dimensional fermion.

 We estimate $\delta m$ by calculating the mass difference between two special states
 $(k+1, \dots , k+1)$  and $(k, \dots, k)$ with masses 
 $m_{k+1} \equiv \sqrt{N}(k+1)/R$  and $m_{k} \equiv \sqrt{N}k/R$, 
 respectively and then dividing their mass difference by the number of states in between the two.

The mass difference between the two special states is  
\begin{equation}
    m_{k+1} - m_{k} = 
    \frac{\sqrt{N}}{R}.
\end{equation}

Now, we estimate the number of different levels in between them (i.e., ignoring the 
level degeneracies). 
That is, we want to find the number of levels that fulfill
\begin{equation}
   m_{k} < m_{k_1,...,k_N} < m_{k+1}.
\end{equation}
Using the expression (\ref{masssplitting}) for the mass levels in the KK tower,
this reduces to
\begin{equation}
    N k^2 < k_1^2 + \dots +k_N^2 < N (k+1)^2.
\end{equation}
For $N>3$, we can express every integer as a sum of integer squares (This is actually a theorem: Lagrange theorem, see, e.g., ~\cite{hardy2008introduction}.)
This is also very accurate for $N=3$, where we can express almost every integer as a sum of squares, as well as for $N=2$ up to a log-factor~\cite{landau1909handbuch}.  
Therefore, in all cases of our interest, the number of levels is essentially given by  the number of integers in between the mass states
\begin{equation}
  \# {\rm mass-levels} =   N(k+1)^2 - N k^2 \approx 2 N k.
\end{equation}

Thus, the mass splitting of the KK tower is
\begin{equation}
    \delta m = \frac{ m_{k+1} - m_{k}}{\# {\rm mass-levels}} \approx \frac{1}{2 \sqrt{N} R k} = \frac{1}{2R^2m_{k}}.
\end{equation}
Now set $m_k \approx m_n$ and we get 
\begin{equation}\label{APPM}
    \delta m  \approx \frac{1}{2R^2m_{n}}.
\end{equation}
 For $N\leq 3$, the above is essentially exact. For $N=2$
the level spacing is less uniform  but is an excellent 
approximation for averaged splitting.  

Indeed, for $N=2$, the number of states in between $m_{k+1}$ and $m_k$ is lower, because we cannot express every integer as a sum of two integer squares.
The number of integers between $0$ and $x$ that can be expressed as a sum of two integer squares goes as $x/\sqrt{\log x}$~\cite{landau1909handbuch}.
Hence, the number of integers we are looking for between $x_1$ and $x_2$ is proportional to
\begin{equation}
    \frac{ x_1}{\sqrt{\log x_1}} - \frac{ x_2}{\sqrt{\log x_2}}.
    \label{eq:nuberofstates}
\end{equation}
For $x_1 = N k^2$ and $x_2 = N (k+1)^2$ and $k \sim 10^{11}$ (remember $k \sim m_n R/\sqrt{2}$), this number is $0.42 k \approx k/2$.

Thus, for $N=2$, the mass splitting of the tower around $m_k \approx m_n$ is
\begin{equation}\label{eq:TwoDimMassSplitting}
    \delta m  \approx \frac{4}{R^2 m_{n}}.
\end{equation}

For a massive bulk fermion with mass $\mu$, we repeat the calculation with,   
\begin{equation}
    m_k^2 = \mu^2 + \frac{k_1^2}{R_1^2} + \dots + \frac{k_N^2}{R_N^2}.
    \label{masssplittingmu2}
\end{equation}
The mass-splitting between the two special states is
\begin{equation}
 m_{k+1} - m_{k} 
\approx \frac{Nk}{R^2 m_{k}} \,.
  \end{equation}

For $N \geq 3$, the number of states in between the two special states is the same 
as in the massless case,  $\# {\rm states} \approx 2 Nk$.  
Correspondingly, so is the level-splitting, 
\begin{equation} 
    \delta m  \approx \frac{1}{2R^2m_{n}}.
\end{equation}
For $N=2$, the number of states is slightly different due to a smaller gap 
 $k^2 \sim (m_n^2 - \mu^2) R^2/2 \approx {\rm MeV GeV} R^2/2$, which 
 limits the maximal $k$, 
 approximately by $k \sim 10^9$  instead of  $k \sim 10^{11}$ of the massless case. 
 This however is a small difference. 
Using again Eq. \eqref{eq:nuberofstates}, $\# {\rm mass-levels} \approx 0.46 k \approx k/2$, 
we also get approximately the same result as in the massless case, \eqref{eq:TwoDimMassSplitting}.

\subsection{Degeneracy of States}

 Using the expression of KK masses Eq. \eqref{masssplitting},
\begin{equation}
    (m R)^2 = k_1^2 + \dots + k_N^2,
\end{equation}
we map the degeneracy count onto a problem in number theory of counting the 
number of different possibilities, $r_N(n)$, of integers 
$k_1, \dots, k_N$
that satisfy $n=k_1^2 + \dots + k_N^2$, with $n=(mR)^2$.
The averaged number of possibilities $r_N(n)$, will give us the degeneracy of states $Z$.
 
 There is a ``cheap" way of getting the answer by making a continuum 
 approximation.  That is, for $R \rightarrow \infty$, the number of states 
 $n(m)$ with mass $\leq m$ is 
 \begin{equation}
   n(m) =  v_N (m R)^{N} \,,
\end{equation}
where $v_N$ is a volume of an unit $N$-ball. 
The number of states within the interval of masses between $m+\Delta m$ and $m$ then is 
 \begin{equation}
   Z(m) =  N v_N \Delta m \, m^{N-1} R^N \,,
\end{equation}
which for $\Delta m = 1/(2m R^2)$ taken from (\ref{APPM}) gives, 
 \begin{equation} \label{ZAPP}
   Z(m) \approx  \frac{N}{2} \, v_N (m R)^{N-2} \,.
\end{equation}

 For an alternative count, we start by estimating all possibilities to solve the equation $k_1^2 + \dots + k_N^2 \leq x$. This can be solved geometrically, as it is just the volume of a ball,
\begin{equation}
    \sum_{n=0}^{x} r_N(n) = v_N x^{\frac{N}{2}}.
\end{equation}
To estimate the average of $r_N$ at some number $x$, we average the $a$ nearest values of $r_N$
\begin{align}
    Z = &\frac{1}{a} (r_N(x-a+1) + \dots + r_N(x))\\
    =& \frac{v_N}{a} (x^{\frac{N}{2}} - (x-a)^{\frac{N}{2}} )\\
    \approx& \frac{N}{2} v_N x^{\frac{N}{2}-1}.
\end{align}
In our case, we replace $x=(mR)^2$ and find for the degeneracy of states
given by (\ref{ZAPP}).

  \bibliographystyle{utphys}
	\bibliography{refs}				         

\providecommand{\href}[2]{#2}\begingroup\raggedright\begin{thebibliography}{10}

\bibitem{Arkani-Hamed:1998jmv}
N.~Arkani-Hamed, S.~Dimopoulos, and G.~Dvali, ``{The Hierarchy problem and new
  dimensions at a millimeter},''
  \href{http://dx.doi.org/10.1016/S0370-2693(98)00466-3}{{\em Phys. Lett. B}
  {\bfseries 429} (1998) 263--272},
  \href{http://arxiv.org/abs/hep-ph/9803315}{{\ttfamily arXiv:hep-ph/9803315}}.

\bibitem{Arkani-Hamed:1998sfv}
N.~Arkani-Hamed, S.~Dimopoulos, and G.~Dvali, ``{Phenomenology, astrophysics
  and cosmology of theories with submillimeter dimensions and TeV scale quantum
  gravity},'' \href{http://dx.doi.org/10.1103/PhysRevD.59.086004}{{\em Phys.
  Rev. D} {\bfseries 59} (1999) 086004},
  \href{http://arxiv.org/abs/hep-ph/9807344}{{\ttfamily arXiv:hep-ph/9807344}}.

\bibitem{Dvali:2007hz}
G.~Dvali, ``{Black Holes and Large N Species Solution to the Hierarchy
  Problem},'' \href{http://dx.doi.org/10.1002/prop.201000009}{{\em Fortsch.
  Phys.} {\bfseries 58} (2010) 528--536},
  \href{http://arxiv.org/abs/0706.2050}{{\ttfamily arXiv:0706.2050 [hep-th]}}.

\bibitem{Dvali:2007wp}
G.~Dvali and M.~Redi, ``{Black Hole Bound on the Number of Species and Quantum
  Gravity at LHC},'' \href{http://dx.doi.org/10.1103/PhysRevD.77.045027}{{\em
  Phys. Rev. D} {\bfseries 77} (2008) 045027},
  \href{http://arxiv.org/abs/0710.4344}{{\ttfamily arXiv:0710.4344 [hep-th]}}.

\bibitem{Dvali:2009ne}
G.~Dvali and M.~Redi, ``{Phenomenology of $10^{32}$ Dark Sectors},''
  \href{http://dx.doi.org/10.1103/PhysRevD.80.055001}{{\em Phys. Rev. D}
  {\bfseries 80} (2009) 055001},
  \href{http://arxiv.org/abs/0905.1709}{{\ttfamily arXiv:0905.1709 [hep-ph]}}.

\bibitem{Dvali:1996xe}
G.~Dvali and M.~A. Shifman, ``{Domain walls in strongly coupled theories},''
  \href{http://dx.doi.org/10.1016/S0370-2693(97)00131-7}{{\em Phys. Lett. B}
  {\bfseries 396} (1997) 64--69},
  \href{http://arxiv.org/abs/hep-th/9612128}{{\ttfamily arXiv:hep-th/9612128}}.
  [Erratum: Phys.Lett.B 407, 452 (1997)].

\bibitem{Dvali:1999gf}
G.~Dvali and G.~Gabadadze, ``{Nonconservation of global charges in the brane
  universe and baryogenesis},''
  \href{http://dx.doi.org/10.1016/S0370-2693(99)00766-2}{{\em Phys. Lett. B}
  {\bfseries 460} (1999) 47--57},
  \href{http://arxiv.org/abs/hep-ph/9904221}{{\ttfamily arXiv:hep-ph/9904221}}.

\bibitem{Arkani-Hamed:1998wuz}
N.~Arkani-Hamed, S.~Dimopoulos, G.~Dvali, and J.~March-Russell, ``{Neutrino
  masses from large extra dimensions", (Presented at SUSY 1998) },''
  \href{http://dx.doi.org/10.1103/PhysRevD.65.024032}{{\em Phys. Rev. D}
  {\bfseries 65} (2001) 024032},
  \href{http://arxiv.org/abs/hep-ph/9811448}{{\ttfamily arXiv:hep-ph/9811448}}.

\bibitem{Dvali:1999cn}
G.~Dvali and A.~Y. Smirnov, ``{Probing large extra dimensions with
  neutrinos},'' \href{http://dx.doi.org/10.1016/S0550-3213(99)00574-X}{{\em
  Nucl. Phys. B} {\bfseries 563} (1999) 63--81},
  \href{http://arxiv.org/abs/hep-ph/9904211}{{\ttfamily arXiv:hep-ph/9904211}}.

\bibitem{Dienes:1998sb}
K.~R. Dienes, E.~Dudas, and T.~Gherghetta, ``{Neutrino oscillations without
  neutrino masses or heavy mass scales: A Higher dimensional seesaw
  mechanism},'' \href{http://dx.doi.org/10.1016/S0550-3213(99)00377-6}{{\em
  Nucl. Phys. B} {\bfseries 557} (1999) 25},
  \href{http://arxiv.org/abs/hep-ph/9811428}{{\ttfamily arXiv:hep-ph/9811428}}.

\bibitem{Berezhiani:2005hv}
Z.~Berezhiani and L.~Bento, ``{Neutron - mirror neutron oscillations: How fast
  might they be?},''
  \href{http://dx.doi.org/10.1103/PhysRevLett.96.081801}{{\em Phys. Rev. Lett.}
  {\bfseries 96} (2006) 081801},
  \href{http://arxiv.org/abs/hep-ph/0507031}{{\ttfamily arXiv:hep-ph/0507031}}.

\bibitem{Berezhiani:2009ldq}
Z.~Berezhiani, ``{More about neutron - mirror neutron oscillation},''
  \href{http://dx.doi.org/10.1140/epjc/s10052-009-1165-1}{{\em Eur. Phys. J. C}
  {\bfseries 64} (2009) 421--431},
  \href{http://arxiv.org/abs/0804.2088}{{\ttfamily arXiv:0804.2088 [hep-ph]}}.

\bibitem{nEDM:2020ekj}
{\bfseries nEDM} Collaboration, C.~Abel {\em et~al.}, ``{A search for neutron
  to mirror-neutron oscillations using the nEDM apparatus at PSI},''
  \href{http://dx.doi.org/10.1016/j.physletb.2020.135993}{{\em Phys. Lett. B}
  {\bfseries 812} (2021) 135993},
  \href{http://arxiv.org/abs/2009.11046}{{\ttfamily arXiv:2009.11046
  [hep-ph]}}.

\bibitem{Ban:2023cja}
G.~Ban {\em et~al.}, ``{Search for Neutron-to-Hidden-Neutron Oscillations in an
  Ultracold Neutron Beam},''
  \href{http://dx.doi.org/10.1103/PhysRevLett.131.191801}{{\em Phys. Rev.
  Lett.} {\bfseries 131} no.~19, (2023) 191801},
  \href{http://arxiv.org/abs/2303.10507}{{\ttfamily arXiv:2303.10507
  [hep-ph]}}.

\bibitem{ATLAS:2021kxv}
{\bfseries ATLAS} Collaboration, G.~Aad {\em et~al.}, ``{Search for new
  phenomena in events with an energetic jet and missing transverse momentum in
  $pp$ collisions at $\sqrt {s}$ =13 TeV with the ATLAS detector},''
  \href{http://dx.doi.org/10.1103/PhysRevD.103.112006}{{\em Phys. Rev. D}
  {\bfseries 103} no.~11, (2021) 112006},
  \href{http://arxiv.org/abs/2102.10874}{{\ttfamily arXiv:2102.10874
  [hep-ex]}}.

\bibitem{CMS:2021far}
{\bfseries CMS} Collaboration, A.~Tumasyan {\em et~al.}, ``{Search for new
  particles in events with energetic jets and large missing transverse momentum
  in proton-proton collisions at $ \sqrt{s} $ = 13 TeV},''
  \href{http://dx.doi.org/10.1007/JHEP11(2021)153}{{\em JHEP} {\bfseries 11}
  (2021) 153}, \href{http://arxiv.org/abs/2107.13021}{{\ttfamily
  arXiv:2107.13021 [hep-ex]}}.

\bibitem{Antoniadis:1998ig}
I.~Antoniadis, N.~Arkani-Hamed, S.~Dimopoulos, and G.~Dvali, ``{New dimensions
  at a millimeter to a Fermi and superstrings at a TeV},''
  \href{http://dx.doi.org/10.1016/S0370-2693(98)00860-0}{{\em Phys. Lett. B}
  {\bfseries 436} (1998) 257--263},
  \href{http://arxiv.org/abs/hep-ph/9804398}{{\ttfamily arXiv:hep-ph/9804398}}.

\bibitem{ParticleDataGroup:2022pth}
{\bfseries Particle Data Group} Collaboration, R.~L. Workman {\em et~al.},
  ``{Review of Particle Physics},''
  \href{http://dx.doi.org/10.1093/ptep/ptac097}{{\em PTEP} {\bfseries 2022}
  (2022) 083C01}.

\bibitem{Lee:2020zjt}
J.~G. Lee, E.~G. Adelberger, T.~S. Cook, S.~M. Fleischer, and B.~R. Heckel,
  ``{New Test of the Gravitational $1/r^2$ Law at Separations down to 52
  $\mu$m},'' \href{http://dx.doi.org/10.1103/PhysRevLett.124.101101}{{\em Phys.
  Rev. Lett.} {\bfseries 124} no.~10, (2020) 101101},
  \href{http://arxiv.org/abs/2002.11761}{{\ttfamily arXiv:2002.11761
  [hep-ex]}}.

\bibitem{Adelberger:2009zz}
E.~G. Adelberger, J.~H. Gundlach, B.~R. Heckel, S.~Hoedl, and S.~Schlamminger,
  ``{Torsion balance experiments: A low-energy frontier of particle physics},''
  \href{http://dx.doi.org/10.1016/j.ppnp.2008.08.002}{{\em Prog. Part. Nucl.
  Phys.} {\bfseries 62} (2009) 102--134}.

\bibitem{Tan:2016vwu}
W.-H. Tan, S.-Q. Yang, C.-G. Shao, J.~Li, A.-B. Du, B.-F. Zhan, Q.-L. Wang,
  P.-S. Luo, L.-C. Tu, and J.~Luo, ``{New Test of the Gravitational
  Inverse-Square Law at the Submillimeter Range with Dual Modulation and
  Compensation},'' \href{http://dx.doi.org/10.1103/PhysRevLett.116.131101}{{\em
  Phys. Rev. Lett.} {\bfseries 116} no.~13, (2016) 131101}.

\bibitem{Hannestad:2001jv}
S.~Hannestad and G.~Raffelt, ``{New supernova limit on large extra
  dimensions},'' \href{http://dx.doi.org/10.1103/PhysRevLett.87.051301}{{\em
  Phys. Rev. Lett.} {\bfseries 87} (2001) 051301},
  \href{http://arxiv.org/abs/hep-ph/0103201}{{\ttfamily arXiv:hep-ph/0103201}}.

\bibitem{Machado:2011jt}
P.~A.~N. Machado, H.~Nunokawa, and R.~Zukanovich~Funchal, ``{Testing for Large
  Extra Dimensions with Neutrino Oscillations},''
  \href{http://dx.doi.org/10.1103/PhysRevD.84.013003}{{\em Phys. Rev. D}
  {\bfseries 84} (2011) 013003},
  \href{http://arxiv.org/abs/1101.0003}{{\ttfamily arXiv:1101.0003 [hep-ph]}}.

\bibitem{Forero:2022skg}
D.~V. Forero, C.~Giunti, C.~A. Ternes, and O.~Tyagi, ``{Large extra dimensions
  and neutrino experiments},''
  \href{http://dx.doi.org/10.1103/PhysRevD.106.035027}{{\em Phys. Rev. D}
  {\bfseries 106} no.~3, (2022) 035027},
  \href{http://arxiv.org/abs/2207.02790}{{\ttfamily arXiv:2207.02790
  [hep-ph]}}.

\bibitem{KamLAND:2005pen}
{\bfseries KamLAND} Collaboration, T.~Araki {\em et~al.}, ``{Search for the
  invisible decay of neutrons with KamLAND},''
  \href{http://dx.doi.org/10.1103/PhysRevLett.96.101802}{{\em Phys. Rev. Lett.}
  {\bfseries 96} (2006) 101802},
  \href{http://arxiv.org/abs/hep-ex/0512059}{{\ttfamily arXiv:hep-ex/0512059}}.

\bibitem{Learned:1979gp}
J.~Learned, F.~Reines, and A.~Soni, ``{Limits on Nonconservation of Baryon
  Number},'' \href{http://dx.doi.org/10.1103/PhysRevLett.43.907}{{\em Phys.
  Rev. Lett.} {\bfseries 43} (1979) 907}. [Erratum: Phys.Rev.Lett. 43, 1626
  (1979)].

\bibitem{Tiesinga:2021myr}
E.~Tiesinga, P.~J. Mohr, D.~B. Newell, and B.~N. Taylor, ``{CODATA recommended
  values of the fundamental physical constants: 2018*},''
  \href{http://dx.doi.org/10.1103/RevModPhys.93.025010}{{\em Rev. Mod. Phys.}
  {\bfseries 93} no.~2, (2021) 025010}.

\bibitem{Ettengruber:2022pxf}
M.~Ettengruber, ``{Neutrino physics in TeV scale gravity theories},''
  \href{http://dx.doi.org/10.1103/PhysRevD.106.055028}{{\em Phys. Rev. D}
  {\bfseries 106} no.~5, (2022) 055028},
  \href{http://arxiv.org/abs/2206.00034}{{\ttfamily arXiv:2206.00034
  [hep-ph]}}.

\bibitem{Dvali:2009fw}
G.~Dvali, I.~Sawicki, and A.~Vikman, ``{Dark Matter via Many Copies of the
  Standard Model},''
  \href{http://dx.doi.org/10.1088/1475-7516/2009/08/009}{{\em JCAP} {\bfseries
  08} (2009) 009}, \href{http://arxiv.org/abs/0903.0660}{{\ttfamily
  arXiv:0903.0660 [hep-th]}}.

\bibitem{Arkani-Hamed:1999rvc}
N.~Arkani-Hamed, S.~Dimopoulos, G.~Dvali, and N.~Kaloper, ``{Many fold
  universe},'' \href{http://dx.doi.org/10.1088/1126-6708/2000/12/010}{{\em
  JHEP} {\bfseries 12} (2000) 010},
  \href{http://arxiv.org/abs/hep-ph/9911386}{{\ttfamily arXiv:hep-ph/9911386}}.

\bibitem{UCNt:2021pcg}
{\bfseries UCN\ensuremath{\tau}} Collaboration, F.~M. Gonzalez {\em et~al.},
  ``{Improved neutron lifetime measurement with UCN$\tau$},''
  \href{http://dx.doi.org/10.1103/PhysRevLett.127.162501}{{\em Phys. Rev.
  Lett.} {\bfseries 127} no.~16, (2021) 162501},
  \href{http://arxiv.org/abs/2106.10375}{{\ttfamily arXiv:2106.10375
  [nucl-ex]}}.

\bibitem{Ezhov:2014tna}
V.~F. Ezhov {\em et~al.}, ``{Measurement of the neutron lifetime with
  ultra-cold neutrons stored in a magneto-gravitational trap},''
  \href{http://dx.doi.org/10.1134/S0021364018110024}{{\em JETP Lett.}
  {\bfseries 107} no.~11, (2018) 671--675},
  \href{http://arxiv.org/abs/1412.7434}{{\ttfamily arXiv:1412.7434 [nucl-ex]}}.

\bibitem{Yue:2013qrc}
A.~T. Yue, M.~S. Dewey, D.~M. Gilliam, G.~L. Greene, A.~B. Laptev, J.~S. Nico,
  W.~M. Snow, and F.~E. Wietfeldt, ``{Improved Determination of the Neutron
  Lifetime},'' \href{http://dx.doi.org/10.1103/PhysRevLett.111.222501}{{\em
  Phys. Rev. Lett.} {\bfseries 111} no.~22, (2013) 222501},
  \href{http://arxiv.org/abs/1309.2623}{{\ttfamily arXiv:1309.2623 [nucl-ex]}}.

\bibitem{Nico:2004ie}
J.~S. Nico {\em et~al.}, ``{Measurement of the neutron lifetime by counting
  trapped protons in a cold neutron beam},''
  \href{http://dx.doi.org/10.1103/PhysRevC.71.055502}{{\em Phys. Rev. C}
  {\bfseries 71} (2005) 055502},
  \href{http://arxiv.org/abs/nucl-ex/0411041}{{\ttfamily
  arXiv:nucl-ex/0411041}}.

\bibitem{Berezhiani:2018eds}
Z.~Berezhiani, ``{Neutron lifetime puzzle and neutron\textendash{}mirror
  neutron oscillation},''
  \href{http://dx.doi.org/10.1140/epjc/s10052-019-6995-x}{{\em Eur. Phys. J. C}
  {\bfseries 79} no.~6, (2019) 484},
  \href{http://arxiv.org/abs/1807.07906}{{\ttfamily arXiv:1807.07906
  [hep-ph]}}.

\bibitem{Tan:2019mrj}
W.~Tan, ``{Neutron oscillations for solving neutron lifetime and dark matter
  puzzles},'' \href{http://dx.doi.org/10.1016/j.physletb.2019.134921}{{\em
  Phys. Lett. B} {\bfseries 797} (2019) 134921},
  \href{http://arxiv.org/abs/1902.01837}{{\ttfamily arXiv:1902.01837
  [physics.gen-ph]}}.

\bibitem{Fornal:2018eol}
B.~Fornal and B.~Grinstein, ``{Dark Matter Interpretation of the Neutron Decay
  Anomaly},'' \href{http://dx.doi.org/10.1103/PhysRevLett.120.191801}{{\em
  Phys. Rev. Lett.} {\bfseries 120} no.~19, (2018) 191801},
  \href{http://arxiv.org/abs/1801.01124}{{\ttfamily arXiv:1801.01124
  [hep-ph]}}. [Erratum: Phys.Rev.Lett. 124, 219901 (2020)].

\bibitem{Karananas:2018goc}
G.~K. Karananas and A.~Kassiteridis, ``{Small-scale structure from neutron dark
  decay},'' \href{http://dx.doi.org/10.1088/1475-7516/2018/09/036}{{\em JCAP}
  {\bfseries 09} (2018) 036}, \href{http://arxiv.org/abs/1805.03656}{{\ttfamily
  arXiv:1805.03656 [hep-ph]}}.

\bibitem{Czarnecki:2018okw}
A.~Czarnecki, W.~J. Marciano, and A.~Sirlin, ``{Neutron Lifetime and Axial
  Coupling Connection},''
  \href{http://dx.doi.org/10.1103/PhysRevLett.120.202002}{{\em Phys. Rev.
  Lett.} {\bfseries 120} no.~20, (2018) 202002},
  \href{http://arxiv.org/abs/1802.01804}{{\ttfamily arXiv:1802.01804
  [hep-ph]}}.

\bibitem{Gonzalo:2022jac}
E.~Gonzalo, M.~Montero, G.~Obied, and C.~Vafa, ``{Dark dimension gravitons as
  dark matter},'' \href{http://dx.doi.org/10.1007/JHEP11(2023)109}{{\em JHEP}
  {\bfseries 11} (2023) 109}, \href{http://arxiv.org/abs/2209.09249}{{\ttfamily
  arXiv:2209.09249 [hep-ph]}}.

\bibitem{hardy2008introduction}
G.~Hardy, E.~Wright, D.~Heath-Brown, and J.~Silverman, {\em An Introduction to
  the Theory of Numbers}.
\newblock Oxford mathematics. OUP Oxford, 2008.
\newblock \url{https://books.google.de/books?id=P6uTBqOa3T4C}.

\bibitem{landau1909handbuch}
E.~Landau, {\em Handbuch der lehre von der verteilung der primzahlen}.
\newblock No.~Bd. 1 in Handbuch der lehre von der verteilung der primzahlen. B.
  G. Teubner, 1909.
\newblock \url{https://books.google.de/books?id=81m4AAAAIAAJ}.

\end{thebibliography}\endgroup

\end{document}